\documentclass[aps,prb,twocolumn,showpacs,amsmath,amssymb,superscriptaddress,10pt,floatfix]{revtex4-2}
\usepackage{graphicx} 
\usepackage[utf8]{inputenc}
\usepackage[LY1]{fontenc}

\usepackage{bm}
\usepackage{dsfont}

\usepackage{orcidlink}
\usepackage{soul}               
\usepackage{xcolor}             

\makeatletter
\newsavebox{\@brx}
\newcommand{\llangle}[1][]{\savebox{\@brx}{\(\m@th{#1\langle}\)}%
  \mathopen{\copy\@brx\kern-0.5\wd\@brx\usebox{\@brx}}}
\newcommand{\rrangle}[1][]{\savebox{\@brx}{\(\m@th{#1\rangle}\)}%
  \mathclose{\copy\@brx\kern-0.5\wd\@brx\usebox{\@brx}}}

\makeatother

\DeclareMathOperator{\Tr}{Tr}

\begin{document}

\title{Predicting the Slow Drift of Nuclear Magnetic Noise in Semiconductor Spin Qubits}
\author{Wayne M. \surname{Witzel}\orcidlink{0000-0002-9082-8076}}
\affiliation{Center for Computing Research, Sandia National Laboratories, 
Albuquerque, New Mexico 87185 USA}
\author{Jesse J. \surname{Lutz}\orcidlink{0000-0003-0296-405X}}
\affiliation{Center for Computing Research, Sandia National Laboratories, 
Albuquerque, New Mexico 87185 USA}
\author{Matthew D. \surname{Grace}\orcidlink{0000-0001-5842-7347}}
\affiliation{Center for Computing Research, Sandia National Laboratories, 
Albuquerque, New Mexico 87185 USA}
\author{Natalie D. \surname{Foster}\orcidlink{0000-0002-3571-2054}}
\affiliation{Sandia National Laboratories, Albuquerque, New Mexico 87185 USA}
\author{Ryan M. \surname{Jock}\orcidlink{0000-0002-1352-0190}}
\affiliation{Sandia National Laboratories, Albuquerque, New Mexico 87185 USA}
\author{Dwight R. \surname{Luhman}\orcidlink{0000-0001-8045-1132}}
\affiliation{Sandia National Laboratories, Albuquerque, New Mexico 87185 USA}

\date{\today} 

\begin{abstract}
The dynamics of a nuclear spin bath generates magnetic noise that is a key contributor to the decoherence of electron spin qubits in electrostatically-defined quantum dots. In this paper, we extend the cluster correlation expansion (CCE) technique, which has proven useful for predicting solid-state qubit coherence times across various settings but is limited to shorter time scales, to incorporate stochastic treatments of cluster dynamics in order to efficiently predict slow drifting Overhauser fields over longer time scales.
This approach combines quantum evolution with classical rate matrices to enable simulation across a wide range of temporal regimes
required to simulate, for example, the long-time convergence of the ergodic $T_2^*$ from Ramsey experiments. 
Our methodology is validated against experimental data from various silicon spin qubit systems, demonstrating a strong agreement between simulation and measurement of Ramsey experiments presented in the form of $T_2^*$ versus averaging time, autocorrelation functions, as well as power spectral densities.  Furthermore, we demonstrate significant back-action effects through modeling and experiment; specifically, the dynamics of the nuclear spin bath depends upon the electron spin occupation schedule.
Finally, our modeling quantitatively predicts the benefits from compensating for the slow drift of Overhauser fields in qubit operations.
Our findings indicate that compensating for an Overhauser rotation measured $\Delta t$ in the past results in an effective $T_2^*$, which we denote $\tilde{T}_2^*(\Delta t)$ for clarity, under certain scenarios of interest, can be one or two orders of magnitude larger than the ergodic $T_2^*$ if the Overhauser rotation is re-characterized every 100 milliseconds; that is, $\tilde{T}_2^*(\Delta t = 100~{\rm ms})$ can be $10$ to $100$ times larger than $T_2^*$. 
This work enhances our understanding of qubit dynamics and nuclear spin noise mitigation.
\end{abstract}

\maketitle

\section{Introduction}

Two important metrics for characterizing the coherence of a qubit are $T_2^*$ and $T_2$.  The $T_2^*$ time is the decay time in a Ramsey experiment which quantifies how quickly different qubits lose phase coherence relative to each other (also known as inhomogeneous broadening).  The $T_2$ time is the decay time in a spin echo experiment (a Hahn echo by default or any other dynamical decoupling sequence of qubit rotations if indicated) which characterizes the timescale for irrecoverable loss of coherence.  The cluster correlation expansion (CCE)
~\cite{Witzel2005,Witzel2006,Yao2006,Witzel2007_deco,Yang2008,Yang2009,Cywiski2009prl,Cywiski2009prb,Witzel2012,Witzel2014}
has proven exceptionally reliable for reproducing spin echo decays (and thereby $T_2$) of solid state spin qubits whose irrecoverable decoherence is dominated by slowly evolving nuclear spins~\cite{Witzel2010,Witzel2007_deco,Cywiski2009prl,Cywiski2009prb,Bluhm2010,George2010,Balian2012}. Additionally, CCE has been instrumental
in the design and evaluation of dynamical decoupling strategies beyond the simple Hahn echo \cite{Witzel2007_deco,Yao2007,Witzel2007_conc,Lee2008,Zhao2011,Zhao2012}.  In this paper, we present an adaptation of CCE that enables computations of a time-dependent value $T_2^*(t_{\rm avg})$, where $t_{\rm avg}$ represents the averaging time for a measurement of $T_2^*$ that converges to the the standard (ergodic) $T_2^*$ in the long time limit [i.e., $T_2^* = \lim_{t_{\rm avg} \to \infty} T_2^*(t_{\rm avg})$].

To realize universal gate operations, both the $T_2^*$ and $T_2$ times must have sufficient duration.
The timescale for reliably storing quantum information 
is determined by $T_2$, which has been demonstrated to
last for about 1 second in donor electron spin qubits in enriched silicon \cite{
Tyryshkin2011,
Muhonen2014}
and for tens of milliseconds for singlet-triplet spin qubits~\cite{
Veldhorst2014,
Yoneda2017}. 
Additionally, a long $T_2^*$ relative to gate times is typically required to achieve high-fidelity gate operations~\cite{Takeda2020,Vahapoglu2022}, except when a dynamically corrected gate strategy~\cite{Khodjasteh2009a} can be employed, which is not always possible~\cite{Khodjasteh2009b}.
In the worst-case scenario, high fidelity gate operations require gates to operate quickly compared with the ergodic $T_2^*$. 
In solid-state qubits, the time dependence of $T_2^*$ is typically determined by the polarization of nuclear spins that induce a slowly drifting Overhauser field.  With our adaptation of CCE, we aim to develop a computational tool for guiding experiments in the optimization of their Overhauser autocorrelation function decay, which directly determines $T_2^*(t_{\rm avg})$, where $T_2^*(t_{\rm avg})$ is defined as the observed $T_2^*$ as a function of averaging time determined by averaging rates ($1/T_2^*$) in quadrature [Appendix~\ref{appendix:T2star}].

When applied to the long timescales of Overhauser autocorrelation function decay, cluster expansions are inaccurate at low orders (small cluster sizes) and prohibitively expensive at high orders (large cluster sizes). 
It can also be crucial to account for quantum back-action effects on a nuclear spin bath as it interacts with different electrons over time~\cite{Fink2014,Bethke2020,Monir2023}.
In general, the bath dynamics are not independent of the electron spins that interact with it.  
We consider the scenario in which a fresh electron is periodically introduced into a quantum dot with a random spin polarization, inducing decoherence of the nuclear spin bath. As phase coherence between different nuclear bath states is lost over time due to interactions with many randomly polarized electron spins, this latter consideration (electron back-action) provides a solution to the former concern (convergence of the cluster expansion). That is because a stochastic influence will induce decoherence of the spin bath itself, which is an effect well approximated using a classical rate-matrix model. 

In the technique that we present in this work, we combine both quantum dynamics and classical rate matrices in a holistic model that can feasibly bridge the coherent short-time behavior and incoherent long-time behavior of a nuclear spin bath therein simultaneously spanning all temporal regimes of interest. We use this technique to calculate $T_2^*(t_{\rm avg})$ for several experimentally relevant scenarios involving Loss-Divincenzo qubits and singlet-triplet qubits. We find good agreement between the simulations and experiments. In addition, we consider the case where regular characterization of individual qubit rotation speeds can be used as real time feedback to extend the effective value of $T_2^*$.

The remainder of the paper is organized as follows: We formulate the basic problem in Sec.~\ref{sec:formulation}; we describe our cluster expansion method for computing Overhauser autocorrelation functions in Sec.~\ref{sec:method}; we make comparisons with available experimental data from Ramsey experiments in Sec.~\ref{sec:experiments} presented as
$T_2^*(t_{\rm avg})$ values, autocorrelation functions, and
power spectral densities (PSDs);
we predict drift-compensated $T_2^*(\Delta t)$, appropriate when correcting for an Overhauser rotation measured in $\Delta t$ time in the past, as a function of $^{29}$Si enrichment and other parameters in Sec.~\ref{sec:predictions}; we conclude in Sec.~\ref{sec:conclusion}, discussing the feasibility of nuclear magnetic noise drift compensation specifically in the context of $^{119}$Sn qubits in silicon \cite{Witzel2022}.

\section{Problem Formulation}
\label{sec:formulation}

Our aim is to calculate the expected time correlation of unknown rotations experienced by a qubit due to the Overhauser field.
The techniques we present are general and can be readily applied to multiple qubit scenarios, as we demonstrate in Sec.~\ref{sec:experiments} when making comparison with experiments.

An electron spin interacts with nuclear spins predominantly through the isotropic Fermi contact hyperfine interaction (HFI),\footnote{We do not expect the anisotropic hyperfine interaction to contribute significantly for silicon quantum dots, in contrast to Refs.~\cite{Witzel2007_deco} where it was demonstrated to be competitive.} while the nuclear spins will interact with each other primarily through the magnetic dipole interaction. Additionally, all of the spins will experience a spin-dependent Zeeman energy shift from an external magnetic field. 
For simplicity, we neglect the time it takes to move an electron on or off a specific quantum dot, which is typically fast (ns scale) relative to nuclear spin dynamics ($\mu$s scale), and toggle between the following two Hamiltonians:
\begin{eqnarray}
\label{eq:Hamiltonians}
{\cal H}_{\rm{off}} &=& {\cal H}_{\rm z} + {\cal H}_{\rm d} \nonumber \\
{\cal H}_{\rm{on}} &=& {\cal H}_{\rm z} + {\cal H}_{\rm{hf}} + {\cal H}_{\rm d},
\end{eqnarray}
where the subscripts `on' and `off' denote whether or not an electron is present in the quantum dot.
The Hamiltonian for the Zeeman effect, ${\cal H}_{\mathrm Z}$, is given by: 
\begin{equation}
{\cal H}_{\mathrm Z} = \gamma_S B \hat{S}_{\mathrm z} - B \sum_n \gamma_n \hat{I}_{n\mathrm z}
\end{equation}
where $\gamma_S$ is the electron gyromagnetic ratio and $\hat{\bm S}$ is the electron spin operator, $\gamma_n$ is the gyromagnetic ratio for the $n$th nuclear spin and $\hat{\bm I}$ is its nuclear spin operator, and the external magnetic field is along the z direction with strength $B$.
The Fermi contact HFI term, ${\cal H}_{\mathrm{hf}}$, is expressed as:
\begin{equation}
\label{eq:hf}
{\cal H}_{\mathrm{hf}} = \hat{\bm S} \cdot \sum_{n} A_n \hat{\bm I}_n \approx \hat{S}_{\mathrm z} \sum_{n} A_n \hat{I}_{n\mathrm z},
\end{equation}
where $A_n = \frac{8 \pi}{3} \gamma_S \gamma_n \lvert \psi_e({\bm r}_n)\rvert^2$ with $\lvert \psi_e({\bm r}_n)\rvert^2$
the probability density of the electronic wavefunction, $\psi_e$, at a given nuclear position, ${\bf r}_n$, when it is loaded on the quantum dot.
This approximation is justified in a modest magnetic field (a few mT) because the electron-spin gyromagnetic ratio is approximately 3 orders of magnitude larger than that of the nuclei ($^{29}$Si, $^{73}$Ge, $^{119}$Sn, and any other common spinful impurities). Therefore any flip-flop between the electron and nuclear spin will be greatly suppressed by the Zeeman energy splitting (see, e.g., Ref.~\cite{Witzel2022}). We also assume low temperature (sub-Kelvin) such that phonon-induced decay is negligible over $\mu$s time scales~\cite{Tyryshkin2003}.
The Hamiltonian for the dipolar interaction, ${\cal H}_{\mathrm d}$, is given by:
\begin{equation}
{\cal H}_{\mathrm d} = \sum_{m < n} \gamma_m \gamma_n \hat{{\bm I}}_m {\mathbb D}({\bm r}_n - {\bm r}_m) \hat{{\bm I}}_n
\end{equation}
where ${\mathbb D}(\bm{r})$ is a tensor characterizing the dipolar interactions, with elements defined as:
\begin{equation}
{\mathbb D}_{\alpha, \beta}({\bf r}) = \left[
\frac{\delta_{\alpha, \beta} - 3 r_{\alpha} r_{\beta}/r^2}{r^3}\right],
\end{equation}
with $\alpha, \beta \in \{x, y, z\}$, $\delta_{\alpha, \beta}$ the Kronecker delta, $r_{\alpha}$ the $\alpha$ component of ${\bm r}$, and $r$ the vector magnitude of ${\bm r}$.

In this paper, we focus on the dynamics of the nuclear spin bath and its specific effect on an electron spin.  The electron may experience an additional spin-orbit interaction~\cite{Jock2018, Ferdous2018_prb}, but this is irrelevant to the magnetic drift considerations discussed here. Under these assumptions, the density matrix of the nuclear spin bath, initially denoted as $\rho_0$, after evolving under ${\cal H}_{\rm off}$ for a time $t_{\rm off}$ and then under ${\cal H}_{\rm on}$ for a time $t_{\rm on}$, will become
\begin{eqnarray}
\nonumber
\rho' &=& \Tr_e{\left[
e^{-i {\cal H}_{\rm on} t_{\rm on}} \left(
e^{-i {\cal H}_{\rm off} t_{\rm off}} \rho_0 e^{i {\cal H}_{\rm off} t_{\rm off}} \otimes \rho_{0e}
\right)
e^{i {\cal H}_{\rm on} t_{\rm on}}
\right]} \\
\nonumber
&=& \frac{1}{2} 
\left(
e^{-i {\cal H}_{\rm on}^{\uparrow} t_{\rm on}} 
e^{-i {\cal H}_{\rm off} t_{\rm off}} \rho_0 e^{i {\cal H}_{\rm off} t_{\rm off}}
e^{i  {\cal H}_{\rm on}^{\uparrow} t_{\rm on}} \right. \\
\label{eq:rho_prime}
&& \left. {} + e^{-i {\cal H}_{\rm on}^{\downarrow} t_{\rm on}} 
e^{-i {\cal H}_{\rm off} t_{\rm off}} \rho_0 e^{i {\cal H}_{\rm off} t_{\rm off}} 
e^{i  {\cal H}_{\rm on}^{\downarrow} t_{\rm on}}
\right),
\end{eqnarray}
where ${\cal H}_{\rm on}^{\uparrow/\downarrow} = {\cal H}_{\mathrm z} + {\cal H}_{\mathrm d} \pm \sum_{n} A_n \hat{I}_{n\mathrm z}/2$ represents ${\cal H}_{\rm on}$ given an up/down electron spin,
and $\rho_{0e}$ is the initial density matrix of an electron that is in an equal superposition of spin-up and spin-down states:
\begin{equation}
\label{eq:rho_0e}
\rho_{0e} = \frac{\left(\lvert \uparrow \rangle + e^{i \phi} \lvert \downarrow \rangle \right) \left(\langle \uparrow \rvert + e^{-i \phi} \langle \downarrow \rvert \right)}{2}.
\end{equation}
Note that the final expression in Eq.~(\ref{eq:rho_prime}) is independent of the initial phase of the electron spin, $\phi$.  This is expected since the nuclear spins should not be sensitive to the electron phase through the $\hat{S}_{\mathrm z} \hat{I}_{n\mathrm z}$ interaction.

For a Ramsey experiment, fresh electron(s) will be introduced to interact with nuclei in the quantum dot(s) for a fixed amount of time and then measured; this may be a single electron probing a single quantum dot or two electrons probing a double quantum dot.  In the future context of running a quantum computation, fresh electrons will be introduced and measured on a regular basis. For the simplicity of our model, we assume an indefinitely repeating pattern of an electron off for a duration of $t_{\rm off}$ and then on for a duration of $t_{\rm on}$ with a periodicity of $T=t_{\rm on}+t_{\rm off}$. A new electron is used in each period. Therefore, Eq.~(\ref{eq:rho_prime}) captures the evolution of the density matrix for any given period, where $\rho_0$ is the state of the nuclear spin bath at the beginning of the period and $\rho'$ is the state after the period. While the details of a particular experiment or setting may deviate from such strict periodicity, exploring the space under this simplification provides useful insights and guidelines and can serve as a reasonable approximation, as we will see in Sec.~\ref{sec:experiments}.

Regarding the HFI, $A_n \propto \lvert \psi_e({\bm r}_n)\rvert^2$, our default, simple model for the electronic density, $|\psi_e({\bm r}_n)\rvert^2$, employs the same proxy wavefunction $\psi_e({\bm r}_n)$ as Eq.~(5) of Ref.~\cite{Witzel2022}, which is based on an infinite square-well potential confining the electron vertically and a parabolic-well potential confining the electron laterally.
At each nuclear site, $n$, characterized by a bunching factor of $\eta_n$, our proxy wavefunction, parameterized by a radius $r_0$ and thickness $z_0$, is given by:
\begin{eqnarray}
    \nonumber
    |\psi_e(x_n, y_n, z_n)|^2 &\propto&
        \eta_n e^{-((x_n-x_0)^2 + (y_n-y_0)^2) / r_0^2}
        \cos^2{\left(\frac{z_n \pi}{z_0}\right)} \\
    & & \times \cos^2{\left(k_0 z_n - \theta_{\mathrm v} / 2\right)},
    \label{eq:wavefunction}
\end{eqnarray}
where $\theta_{\mathrm v}$ is the valley phase (assuming all but the two vertically oriented valley states in the six-fold degenerate conduction-band minima of bulk silicon are split far away in energy due to the vertical confinement of the quantum well).  
The valley oscillation frequency is based on effective mass theory for silicon,
$k_0 = 0.85 \cdot 2 \pi / a_0$ with $a_0 = 0.543$~nm as a standard silicon lattice constant~\cite{Tiesinga2021}. 
The quantum dot radius is characterized by $r_0$, and the thickness is characterized by $z_0$ in this model.
For $^{29}$Si and $^{73}$Ge, we use bunching factors of $\eta_{\mathrm{Si}} = 178$~\cite{Assali2011} and $\eta_{\mathrm{Ge}} = 570$~\cite{Kerckhoff2021}, respectively.

For a MOS quantum dot where an electron is vertically confined by an electric field that pulls the electron against an oxide interface, an Airy function is more appropriate than a sinusoid in the $z$ direction.  For a Si/SiGe heterostructure, the wavefunction can be approximated by solving a 1-D Schr\"odinger equation with the potential energy proportional to the vertical profile of the alloy composition \cite{Beddard2011}. The vertical profile of the quantum-dot, which was approximated as a logistic sigmoidal function representing the $\tau$-dependent Si/SiGe profile \cite{Dyck2017}, was taken to be 180~meV deep throughout this work, following  Ref.~\cite{Schffler1997}. 
Finer details about the quantum-dot wavefunction are unlikely to be relevant for our purposes.
Each of our results will indicate the model that we employed.

\section{Method}
\label{sec:method}

When a bath contains many tens to hundreds of nuclear spins or more, exact numerical dynamics simulations
become infeasible. A combination of refactorization and approximation can reduce the computational scaling with system size. In this context, cluster expansions~\cite{Yang2008} are a particularly
valuable approach, but low-order approximations are effective only within a limited temporal range. 
Cluster expansion techniques have been applied to various problems in solid-state
physics to simulate the effects of a bath of many nuclear spins on a single central electronic spin 
\cite{Yang2016}. 
They are also highly adaptable and can take various forms depending on the choice of physical observable 
and series expansion. Such expansions are typically designed 
to converge to the exact value of the quantum observable in the limit of including the full hierarchy of cluster sizes.
By limiting the size of contributing clusters, one may achieve a reasonable approximation in the short-time regime that remains computationally feasible.

In practice, low-order approximations are often sufficient for capturing short-time behavior
because, in an appropriate setting, larger cluster contributions become appreciable only on 
longer time scales.
In the long-time regime, however, the Overhauser field can be
simulated as a classical random telegraph noise (RTN) process~\cite{vanKampen1992}. 
Within this
model, $I=1/2$ bath spins are treated as two-level fluctuators (TLFs): independent, bistable systems
that transition at random intervals (for additional properties of TLFs, see Appendix \ref{appendix:classical_PSD}). 
The objective of the present study is to combine a TLF
model -- and more generally, for spins with $I>\frac{1}{2}$,
a multi-level fluctuator model -- with a cluster expansion to achieve a holistic method 
that can accurately treat all pertinent time regimes.

In Sec.\ \ref{sec:cce}, we review the CCE technique, 
while in Secs.~\ref{sec:avging}--\ref{sec:classical} we move to present
our modifications that enable accurate estimation of
long-time behavior, while maintaining the good accuracy in the short-time regime as established previously.
Our current development of the theory builds upon the CCE formulation of Ref.~\cite{Witzel2014} 
adapted to the context of the problem described in Sec.~\ref{sec:formulation}.  
To the best of our knowledge, the methods described in Secs.~\ref{sec:avging}--\ref{sec:classical}
are presented here for the first time.

\subsection{Cluster expansion framework}
\label{sec:cce}

The short-time behavior of a nuclear spin bath is well-approximated using a cluster expansion technique \cite{Yang2008}. 
A cluster expansion of the bath interactions involves re-factorization, by cluster size,
of all correlation contributions to a quantum observable. At the lowest order, each spin in the bath is
treated as an independent contributor (i.e., no many-body interactions are included).
The first order is trivial, as it describes only contributions to the 1-cluster dynamics
attributable to the Zeeman interaction from a static magnetic field.
At second order, the dynamics of flip-flopping bath spin pairs are included, where, in the simplest form of CCE, the
pairs are treated as evolving independently of any neighboring spin flips.

The CCE approach developed in Ref.\ \cite{Witzel2014} differs from other CCE formulations (see, e.g., 
Refs.\ \cite{Witzel2005,Witzel2006,Yao2006,Yang2008,Witzel2012}) 
in that it targets the autocorrelation function of the bath as the quantum observable of interest, 
instead of a quantum observable related to a fidelity. In making this transition,
the CCE formalism was adapted so that clusters made additive contributions to the relative 
autocorrelation function (which is zero at the initial time) 
instead of multiplicative contributions to a fidelity (with a value of one at the initial time).  
Both approaches provide an accurate approximation in the
short-time regime and an exact result in the limit of including all clusters (apart from
division-by-zero issues that may arise in the multiplicative version).
By turning the focus to estimation of the autocorrelation function, 
one is able to study properties of the spin bath directly 
rather than focusing solely on the bath's influence on a central spin.

We start by defining an initial state of the nuclear bath, $\rho_0$,
within an $N$-dimensional Hilbert space, and an autocorrelation function, $R(t)$,
of the Overhauser field, $\hat{\Omega} = \sum_{n} A_n \hat{I}_{n\mathrm{z}}/2$, defined as:
\begin{equation}
\label{eq:RtDef}
R_{\rho_0}(t) = \langle \hat{\Omega}(0) \hat{\Omega}(t) \rangle_{\rho_0} = \Tr{\left[\hat{\Omega}(0) \hat{\Omega}(t) \rho_0 \right] },
\end{equation}
where $\hat{\Omega}(t)$ indicates the operator is represented in the Heisenberg picture. Here we assume 
that $\rho_0$ is diagonal in the basis of $\hat{I}_{n\mathrm{z}}$ so that
\begin{equation}
\label{eq:RtExplicit}
R_{\rho_0}(t) =
\sum_k p_k R_{\rho_k}(t)
= \sum_k p_k \hat{\Omega}_{kk} \Tr{\left[\hat{\Omega}(t) \rho_k \right] }
\end{equation}
where $p_k$ is the initial occupation probability of each $\rho_k$ state, such that
$\rho_0 = \sum_k p_k \rho_k$, $\hat{\Omega}_{kk} = \Tr{[\hat{\Omega} \rho_k]}$, and 
$\rho_k = \lvert \Psi_k \rangle \langle \Psi_k \rvert$ with 
each $\lvert \Psi_k \rangle$ being an eigenstate of 
all $\hat{I}_{n\mathrm{z}}$.  By virtue of $\hat{\Omega}(t)$ being an operator in the Heisenberg picture,
\begin{eqnarray}
\nonumber
\Tr{\left[\hat{\Omega}(t) \rho_k \right]} &=& \Tr\left[\hat{\Omega} \rho_k(t)\right] \\
\lvert \rho_k(t_0 + t) \rrangle &=& {\cal M}(t) \lvert \rho_k(t_0) \rrangle,
\end{eqnarray}
where $t = mT$ and ${\cal M}(t)$ is the dynamical map that transforms the density matrix according to Eq.~\ref{eq:rho_prime}
from a state at time $t_0$ to its state at a time after $m$ periods of $T$. 
Here we introduce ${\cal M}$ as a superoperator that propagates evolution of nuclei having arbitrary spin quantum number. In the $I=\frac{1}{2}$ case, for example, ${\cal M}$ is an $2^N \times 2^N$ matrix, defined by its action on a vectorized density matrix which is denoted as $\lvert \rho \rrangle$~\cite{Preskill1998,Breuer2002}.

It is advantageous to separate the problem into contributions from different clusters. To do this, we first define $R_{\rho_k}^{\cal S}(t)$
as the autocorrelation function arising from the hypothetical scenario in which all nuclear spins
except those in the cluster ${\cal S}$ are artificially frozen in their initial state according to $\rho_k$.
This may be expressed as:
\begin{eqnarray}
\label{eq:RtS}
\nonumber
R_{\rho_k}^{\cal S}(t) &=& \hat{\Omega}_{kk}
\Tr[\hat{\Omega} \rho_k^{\cal S}(t)] \\
\rho_{k}^{\cal S}(t) &=& \left[{\cal M}^{\cal S}(t)\right] \rho_k(0),
\end{eqnarray}
where the dynamical map ${\cal M}^{{\cal S}}(t)$ is defined according to Eq.~(\ref{eq:rho_prime}) for each period,
except that we exclude all terms in the Hamiltonians [Eq.~\ref{eq:Hamiltonians}] that do not preserve the state of any nuclear spin external to ${\cal S}$.
We define the evolutionary quantities
\begin{eqnarray}
\nonumber
L_k = R_{\rho_k}(t) - R_{\rho_k}(0)~\textrm{and} \\ 
L_{k}^{\cal S} = R_{\rho_k}^{\cal S}(t) - R_{\rho_k}^{\cal S}(0),
\end{eqnarray}
and next we proceed to define $\tilde{L}_{k}^{\cal S}$ inductively via
\begin{eqnarray}
\label{eq:cluster_contribs}
L_{k}^{\cal S} &=& \sum_{{\cal C} \subseteq {\cal S}} \tilde{L}_{k}^{\cal S}~\textrm{which implies} \nonumber \\
\tilde{L}_{k}^{\cal S} &=& L_{k}^{\cal S} - \sum_{{\cal C} \subset {\cal S}} \tilde{L}_{k}^{\cal C}.
\end{eqnarray}
Our approximate expansion, corresponding to a given initial state $\rho_k(t=t_0)$, is ultimately:
\begin{equation}
L_{k} \approx \sum_{C \in \Upsilon} \tilde{L}_{k}^{\cal C}.
\end{equation}
where $\Upsilon$ is the set of clusters defining the approximation.  

In addition to limiting the cluster size, $\Upsilon$ may be restricted to a subset of clusters that exhibit the strongest dipolar interactions among constituents.  In our calculations, we specify a subset of strongly interacting pairs per spin and generate clusters from all possible connected graphs formed by edges defined by these pairs. A specified number of pairs, $k$, defines a graph where each spin is a vertex with a degree of at least $k$, and its edges includes the $k$ spins with the strongest dipolar interactions.  Clusters are formed from the connected subgraphs. One may alternatively define the graph based on a dipolar strength cut-off, but here we opted for specifying the number of pairs $k$ as it has the convenience of being universal across enrichment levels.

Note that $L$ becomes exact, by definition, when the set $\Upsilon$ includes all possible clusters. We define the $l$th
order of the CCE approximation of $L_{k}$ with a maximum cluster size $l$ as:
\begin{equation}
\label{eq:l_cluster_approx}
L_{k}^{(l)} \equiv \sum_{\| C \| \leq l} \tilde{L}_{k}^{\cal C} \approx \sum_{C \in \Upsilon~|~\| C \| \leq k} \tilde{L}_{k}^{\cal C}.
\end{equation}
Similar to choosing $\Upsilon$ to limit which cluster contributions are included, we also use ${\cal M}^{{\cal S}, \Upsilon'}$ in practice instead of ${\cal M}^{\cal S}$ in Eq.~(\ref{eq:RtS}) where ${\cal M}^{{\cal S}, \Upsilon'}$ is defined to only include the Ising interactions between pairs that are contained in $\Upsilon'$.  We typically use $\Upsilon' = \Upsilon$ for simplicity.  This is done for efficiency to avoid the need to include $O(n^2)$ interaction terms given $n$ spins.
 
We estimate $R_{\rho_0}(t)$ with an $l$-cluster approximation by averaging estimates of $L_{k}^{(l)}$ (according to $\Upsilon$ and $\Upsilon'$) using a Monte-Carlo approach where we draw random samples of $\lvert \Psi_k \rangle$ according to $\rho_0$.  In our simulations, we use a uniform distribution, taking $\rho_0$ to be the maximally mixed (infinite temperature) state since typical operating temperatures ($\sim 100$~mK) are high relative to nuclear Zeeman energies ($\sim 1$~nK / mT).
In Sec.~\ref{sec:avging}, we will discuss a further approximation that avoids having to average over instances of $\lvert \Psi_k \rangle$ that is applicable when we can entirely neglect interactions with spins external to a cluster (i.e., $\Upsilon' = \emptyset$).

\subsection{Cluster state averaging with external spin awareness}
\label{sec:avging}

The approximation introduced here as `external spin awareness' (adopting terminology from Ref.~\cite{Witzel2012}) is invoked if and only if $\Upsilon' \neq \emptyset$. When $\Upsilon' = \emptyset$, indicating no external spin awareness, each $\tilde{L}_k^{\cal C}$ is truly independent of any spin state outside of ${\cal C}$.  Therefore, when estimating $R_{\rho_0}(t)$ with an $l$-cluster approximation by averaging estimates of $L_k^{(l)}$ [Eq.~(\ref{eq:l_cluster_approx})] for a $\rho_0$ that is a product state for each nuclear spin, we can swap the order of operations with respect to the summation over ${\cal C}$ and the averaging over $k$. In swapping these operations, we only need to average over the states of the spins within the cluster for each ${\cal C}$.  That is,
\begin{equation}
\sum_{k=1}^N p_k \sum_{\cal C} \tilde{L}_k^{\cal C} =
\sum_{\cal C} \sum_{k' = 1}^{N_{\cal C}} p_{\left[k'~{\rm for}~{\cal C}\right]} \tilde{L}_{\left[k'~{\rm for}~{\cal C}\right]}^{\cal C},
\label{eq:19}
\end{equation}
where the $k'$ summation is only over $N_{\cal C}$ states of the spins of ${\cal C}$ and $\left[k'~{\rm for}~{\cal C}\right]$ indicates that $k'$ only specifies the state of the spins of ${\cal C}$.
We refer to this process as cluster state averaging, and it represents a significant simplification when $\Upsilon' = \emptyset$ (there is no external spin awareness).

When external spin awareness is invoked, we can still employ a form of cluster state averaging when $\rho_0$ is a product state for each nuclear spin.  We cannot eliminate the averaging over states beyond clusters, but 
we can accelerate the convergence of Monte Carlo averaging by also summing over just the cluster states in addition to the external averaging over all states:
\begin{eqnarray}
\label{eq:state_avging_with_esa}
\sum_{k=1}^{N} p_k \sum_{\cal C} \tilde{L}_k^{\cal C} &=&
\left\langle \sum_{\cal C} \tilde{L}_k^{\cal C} \right\rangle_k  \\
\nonumber
&=&
\left\langle \sum_{\cal C} \sum_{k'=1}^{N_{\cal C}} p_{\left[k'~{\rm for}~{\cal C}\right]} \tilde{L}_{\left[k~{\rm except}~k'~{\rm for}~{\cal C} \right]}^{\cal C} \right\rangle_k,
\end{eqnarray}
where $\left[k~{\rm except}~k'~{\rm for}~{\cal C} \right]$ indicates that $k'$ specifies that states to use for the spins within ${\cal C}$ while $k$ specifies the state to be used for all other spins.
This accelerates convergence of Monte-Carlo averaging by 
maintaining equal contributions from each state of a given cluster, which are the most 
significant states when computing a cluster contribution.  
This improves Monte Carlo convergence by reducing the variance across instances of $k$ for each ${\cal C}$.

\subsection{Accounting for flip-flops with spins that are external to a cluster}
\label{sec:effe}
The main contribution of this work is to account for flip-flops involving spins that are external to a cluster in a semi-classical manner, which serves the purpose of improving cluster expansion convergence in the long time regime.  Our procedure is described at a high level as follows:  We pre-compute minimum flip-flop rates (corresponding to probability decays in the long-time limit)
for isolated state transitions between pairs of spins, which are determined by
the dynamical map ${\cal M}^{{\cal S},\Upsilon'}$ where $S$ is the set containing the spins of the pair (or more generally ${\cal M}_{ij}^{\Upsilon'}$ to be described below where $i$ and $j$ denote the transitioning states). We use this information to estimate flip rates for each particular spin, $s \in S$, by summing over contributions from flip-flops of $\{s, x\}$ pairs in $\Upsilon$ such that $x \notin S$.
In the limit of infinite nuclear bath temperature in our simulations, flip-flop contributions from $^{29}$Si pairs are simply the flip-flop rate divided by $2$ to account for the probability that the pair of spins are anti-aligned.  For a bath with a biased polarization (e.g., finite temperature), the flip-flop contributions for flipping up versus down should be scaled separately according to the probability of each $\hat{I}_{nz}$ eigenstate for each partner spin.  This approach can similarly be generalized for spin quantum numbers greater than $1/2$ (e.g., $^{73}$Ge for SiGe-based quantum dots).  We then modify ${\cal M}_{{\cal S}, \Upsilon'}$ to account for the flip rates due to interactions that are external to ${\cal S}$ in a cluster expansion calculation.

Delving into the details of the implementation, we will first describe how flip-flop (transition) rates are determined. Next, we will mathematically express how flip rates are derived from flip-flop rates, and, finally, we will explain how flip rates are incorporated into our cluster simulation.

Let us define $\vec{p}(t)$ as the time dependent occupation of $\hat{I}_{n\mathrm{z}}$ eigenstates (the diagonal elements of $\rho(t)$ in the
$\hat{I}_{n\mathrm{z}}$ basis).  Classical behavior can be described using a rate equation of the form
\begin{equation}
\label{eq:rateEqn}
\frac{\mathrm d}{{\mathrm d}t} \vec{p}(t) \approx \Gamma \vec{p}(t),
\end{equation}
where $\Gamma$ is a linear operator (and $N \times N$ matrix).  
We aim to define $\Gamma$ such that each element corresponds with transition rates of the
quantum dynamics in the long-time limit.  In this way, we approximate the quantum behavior with a classical rate matrix signified by the $\approx$ symbol in Eq.~(\ref{eq:rateEqn}).  To establish this correspondence, we integrate this rate equation to obtain
\begin{equation}
\label{eq:rateEqnIntegrated}
\vec{p}(t) \approx \exp(\Gamma t) \vec{p}(0),
\end{equation}
which is valid under the assumption that $\Gamma$ is diagonalizable.

Conversely, one may approach the same problem using an open quantum systems formulation. In this case, one assumes a time-dependent dynamical map ${\cal M}(t)$ and vectorized density matrix $\lvert \rho(t) \rrangle$. The state of the system evolves according to
\begin{equation}
\lvert \rho(t) \rrangle = {\cal M}(t) \lvert \rho(0) \rrangle = \left[ {\cal M}(T) \right]^m \lvert \rho(0) \rrangle,
\end{equation}
where $t = m T$ and $m$ is the number of periods of duration $T$.  
For simplicity and robustness in our procedure, each transition rate for given states $i$ and $j$ is computed separately based on Hamiltonian operators projected onto these particular states.  That is, we compute ${\cal M}_{ij}^{\Upsilon'}(T)$ much like 
${\cal M}^{{\cal S},\Upsilon'}(T)$, 
where ${\cal S}$ is the set of two spins involved in the transition between $i$ and $j$. However, we project the Hamiltonians of Eq.~(\ref{eq:Hamiltonians}) onto the space of just the $i$ and $j$ states before generating the dynamical map.  The transition rate will be zero for any pair of states that do not directly interact through Hamiltonian terms.

Evolving the density matrix using repetitions of this ${\cal M}_{ij}^{\Upsilon'}$ map and
then projecting this onto classical probabilities associated with the diagonal elements of the density matrix gives
\begin{equation}
\label{eq:projectedDynEqn}
\vec{p}_{ij}(t) \approx P_C \left({\cal M}_{ij}^{\Upsilon'}(T) \right)^m P_C^T \vec{p}_{ij}(0).
\end{equation}
where $P_C$ projects from the vectorized density matrix with four components onto the two component vector of $i/j$ real-valued state probabilities (the C subscript denotes that this is a projection onto the ``Classical" space with no coherent superposition states).  We can obtain the effective contribution to $\Gamma$ according to Eq.~(\ref{eq:rateEqnIntegrated}):
\begin{eqnarray}
\label{eq:gammaApprox}
\Gamma_{ij} &\geq& \lim_{m \rightarrow \infty}
\langle i \vert \ln\left(P_C \left({\cal M}_{ij}^{\Upsilon'}(T) \right)^m P_C^T \right) \vert j \rangle/mT
\end{eqnarray}
for all $i \neq j$.  The diagonal elements of $\Gamma$ are dictated by the constraint that probability must be conserved: $\sum_k \Gamma_{kj} = 0$.

Given the decoherence inherent in ${\cal M}_{ij}$ (except when hyperfine energies are exactly degenerate), the eigenvectors of $P_C \left({\cal M}_{ij}^{\Upsilon'}(T) \right)^m P_C^T$ should converge, in the large $m$ limit, to the maximally mixed state $\left(\lvert i \rangle + \lvert j \rangle\right)/\sqrt{2}$
 and its orthogonal complement $\left(\lvert i \rangle - \lvert j \rangle\right)/\sqrt(2)$.  Since the maximally mixed state is a fixed state of the process, its corresponding eigenvalue will be $1$ and will contribute nothing to the transition rate since $\ln(1) = 0$.  Thus,
\begin{eqnarray}
\label{eq:gammaApproxSimp}
\Gamma_{ij} &\geq& \lim_{m \rightarrow \infty}
\frac{-\ln\left(
\langle v \rvert P_C \left({\cal M}_{ij}^{\Upsilon'}(T) \right)^m P_C^T \lvert v \rangle)\right)}{2mT},
\end{eqnarray}
where $\lvert v \rangle = \left(\lvert i \rangle - \lvert j \rangle \right)/\sqrt{2}$ and
$\langle i \vert v \rangle \langle v \vert j \rangle = -1/2$.

Except, perhaps, for pathological instances, ${\cal M}_{ij}^{\Upsilon'}(T)$ should have an eigendecomposition of the form
\begin{equation}
{\cal M}_{ij}^{\Upsilon'}(T) = Q_{ij} \Lambda_{ij} Q_{ij}^{-1},
\end{equation}
where the $\Upsilon'$ dependence on the right side is implicit and
$\Lambda_{ij}$ is a diagonal matrix composed of the four eigenvalues of ${\cal M}_{ij}^{\Upsilon'}(T)$ which we will denote $\lambda_0$, $\lambda_1$, $\lambda_2$, and $\lambda_3$ sorted firstly according to whether corresponding right/left eigenvectors have a non-negligible overlap with the $P_C$ projection and secondly from largest to smallest in absolute value.  We expect $\lambda_0 = 1$ corresponding with the maximally mixed state as discussed above.  The largest contributor 
to $\Gamma_{ij}$ will be from $\lambda_1$ which may be real-valued or complex and degenerate.  The latter case must be degenerate because the positivity of state probabilities must be maintained (for any $m$) and, for this reason, $\lambda_2 = \lambda_1^*$ unless $\lambda_1$ is real-valued.  Furthermore, the remaining factors for these two contributions must also be complex conjugates of each other.  Thus, in this case,
\begin{eqnarray}
P_C \left({\cal M}_{ij}^{\Upsilon'}(T) \right)^m P_C^T &=&
\lambda_1^m \alpha + (\lambda_1^*)^m \alpha^* \\
\nonumber
&=& |\lambda_1|^m |\alpha| \cos{(m \arg{(\lambda_1)} + \arg{(\alpha)})} \\
\nonumber
&\leq & 2 |\lambda_1|^m |\alpha|.
\end{eqnarray}
and therefore
\begin{equation}
\Gamma_{ij} \geq \lim_{m \rightarrow \infty}
\frac{-\ln{\left(2 |\lambda_1|^m |\alpha|\right)}}{2mT} \geq \frac{-\ln{|\lambda_1|}}{2T},
\end{equation}
This bound applies just as well when $\lambda_1$ is real and non-degenerate since constant factors (independent of $m$) have no effect in the large $m$ limit.

The process above is used to determine $\Gamma_{\{s, x\}, \Upsilon'}$ from ${\cal M}_{\{s, x\}, \Upsilon'}$ for each pair of nuclear spins $\{s, x\}$, contained in $\Upsilon$.  For a given ${\cal S}$, we modify ${\cal M}_{{\cal S}, \Upsilon'}$ to account for flip-flops with external pairs for pairs contained in $\Upsilon$.  
For efficiency, we do this by first computing flip-rates for each spin, $s$, by summing contributions from flip-flop rates with each other spin $x$ for which $\{s, x\} \in \Upsilon$.  Then for a given ${\cal S}$ we subtract contributions from flip-flop rates with each $y \in {\cal S}$ since internal flip-flops are included appropriately in ${\cal M}_{{\cal S}, \Upsilon'}$ already.  The flip-flop rate for a given $\{s, x\}$ pair contributes to the flip rate of $s$, according to
\begin{equation}
\Gamma_{\{s\}, \Upsilon'} = \sum_{x~|~\{s, x\} \in \Upsilon'}
\sum_{k} p_{x, k}
\Tr_2 \left( \Gamma_{\{s, x\}, \Upsilon'} 
\left( I_s \otimes \lvert k \rangle \langle k \rvert \right) \right) \
\end{equation}
where $p_{x, k}$ is the probability for $x$ to be in state $\lvert k \rangle$ according to $\rho_0$, $\Tr_2$ traces over the degrees of freedom of $x$, and $I_s$ is the identity matrix for the $N_{\{s\}}$ states of $s$.

For simplicity, we incorporate these external flip-flop rates into ${\cal M}_{{\cal S}, \Upsilon'}(T)$ with a Trotter-like approximation.  That is, we use
\begin{equation}
\label{eq:dynMapWithExtFlipFlops}
{\cal M}_{{\cal S}, \Upsilon'}'(T) \approx \exp\left({\bf \Gamma}^{\rm ext}_{{\cal S}, \Upsilon', \Upsilon} T\right) {\cal M}_{{\cal S}, \Upsilon'}(T),
\end{equation}
in place of ${\cal M}_{{\cal S}, \Upsilon'}(T)$ in the cluster approximation where ${\cal M}_{{\cal S}, \Upsilon'}(T)$ on the right side is determined from Eq.~{\ref{eq:rho_prime}} and ${\bf \Gamma}^{\rm ext}_{{\cal S}, \Upsilon', \Upsilon}$ is the rate matrix determined from flip-flop rates with external spins as described above as a superoperator given appropriate restrictions from $\Upsilon$ (to limit the pairs contributing flip-flop rates)  and $\Upsilon'$ (to limit inclusion of effective magnetic fields from external spins when calculating the flip-flop rate of a given pair).  Using the above approximation does not change the fact that the cluster expansion is exact in the limit of including all clusters since there is nothing external to a cluster that encompasses everything.

\begin{figure}
\includegraphics[width=\linewidth]{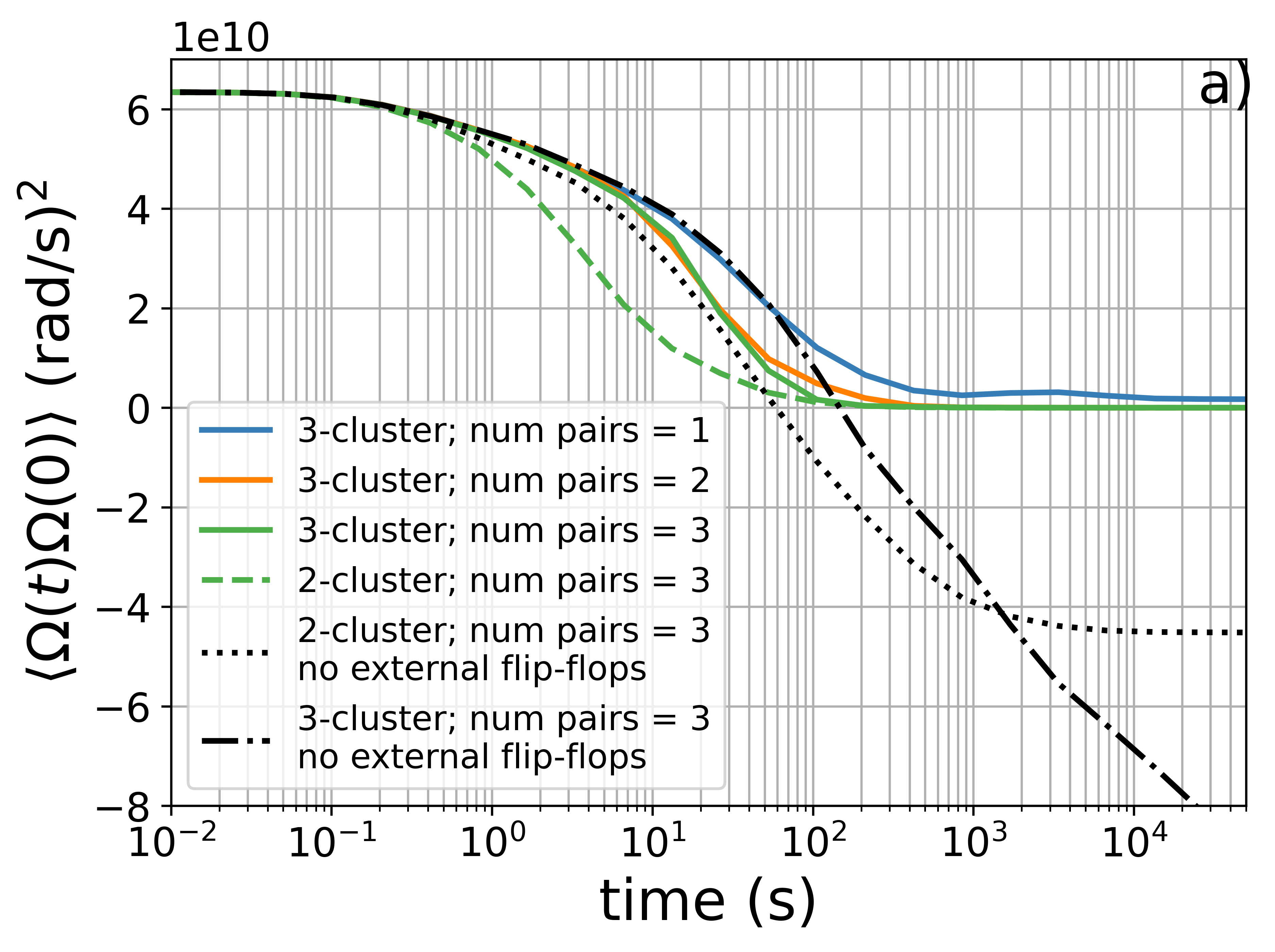} \\
\includegraphics[width=\linewidth]{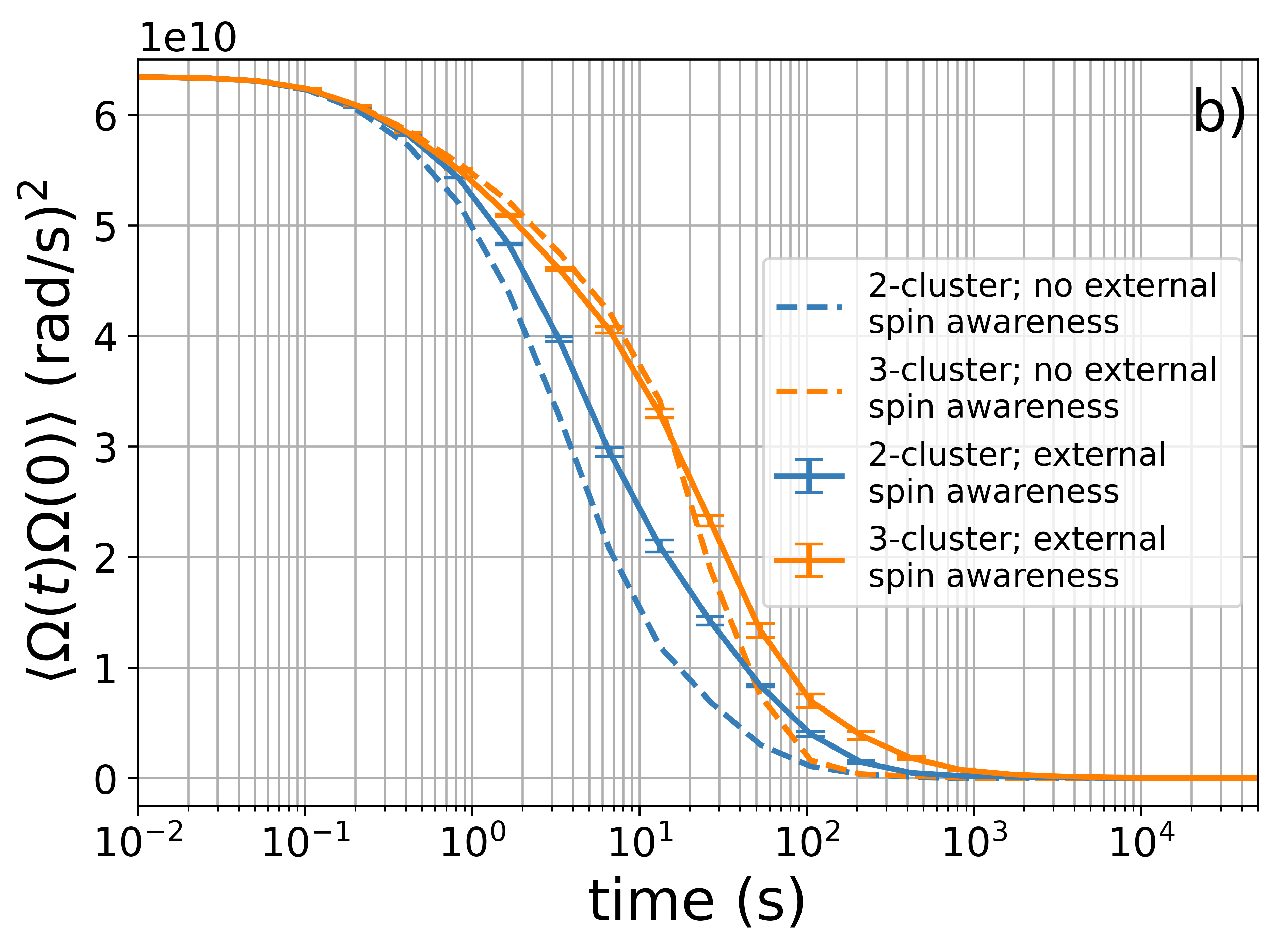} \\
\caption{
\label{fig:ff_comparison}
Autocorrelation functions illustrating the improved convergence behavior when (a) accounting for flip-flops with spins external to each cluster and (b) additionally including external spin awareness ($\Upsilon' = \{C \in \Upsilon~{\rm s.t.}~\|C\|=2\}$).  We utilize full cluster state averaging in all cases; with external spin awareness we use 5 different random spin states (i.e., $k=5$ in Eq.~\ref{eq:state_avging_with_esa}) with the negligibly small error bars showing the standard error of the mean.
The same random nuclear isotope constellation was used throughout, corresponding to 500~ppm $^{29}$Si enrichment, and a quantum dot with $r_0 = $10~nm, $z_0 = 5$~nm, and $\theta_{\mathrm v} = 0$ using the model of Eq.~\ref{eq:wavefunction} and a magnetic field along the [001] crystallographic direction in the secular approximation (infinite magnetic field limit).  
We chose $t_{\rm on} = 1~\mu$s and $t_{\rm off} = 100~\mu$s for these examples.
Panel (a) also demonstrates convergence as we include more pairwise interactions; for each spin, we include pairwise interactions to spins with which it has the strongest dipolar interaction strength up to the specified number of pairs (num pairs).
}
\end{figure}

Fig.\ \ref{fig:ff_comparison} demonstrates the value of accounting for external flip-flops.  It shows calculated autocorrelation functions with/without including external flip-flops when including clusters up to size 2 or 3.  The short-time behavior is consistent for all of these results, but the long-time curves are only well-behaved when external flip-flops are included.  The curves properly approach zero asymptotically at long times only when including external flip-flops.  They also demonstrate convergence as we increase the number of pairwise interactions among bath spins.  The results of the top panel used $\Upsilon' = \emptyset$ while the bottom panel demonstrates that the convergence is further improved with $\Upsilon' = \{C \in \Upsilon~{\rm s.t.}~\|C\|=2\}$ for external spin awareness.  All of these calculations were performed for one random instance of nuclear isotope locations for 500~ppm $^{29}$Si enrichment and a quantum dot with a 10~nm radius and 5~nm thickness.  

\begin{figure}
\includegraphics[width=\linewidth]{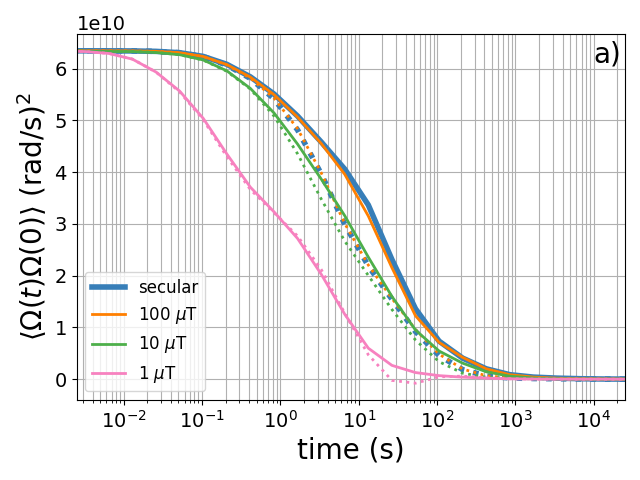}
\includegraphics[width=\linewidth]{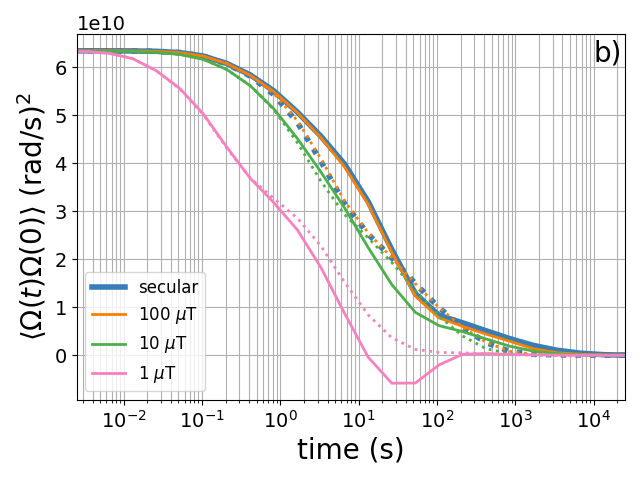}
\caption{
\label{fig:Bfield_comparison}
Autocorrelation functions for different magnetic field strengths up to the secular approximation limit for the same random instance 
of 500~ppm $^{29}$Si, quantum dot shape, $t_{\rm on}$, and $t_{\rm off}$ as in Fig.~\ref{fig:ff_comparison} with the magnetic field along a) [001] and b) [110] crystallographic directions.  Solid (dotted) curves are restricted to clusters of size 3 (2).  
We include external flip-flops and external spin awareness with full cluster state averaging and 5 random spin states as in Fig.~\ref{fig:ff_comparison}~(b).
The secular approximation appears to work well for magnetic fields at 100~$\mu$T or above for 500~ppm $^{29}$Si.
}
\end{figure}

We used a secular approximation that is applicable when the magnetic field is sufficiently large such that the net nuclear spin polarization is preserved.  This approximation is good at 100~$\mu$T or above as demonstrated in Fig.~\ref{fig:Bfield_comparison} where we applied different B-field orientations and magnitudes to this same random instance of $^{29}$Si locations.

\subsection{Classical approximation}
\label{sec:classical}

In the previous section, we showed how we can incorporate effects of stochastic flip-flops with external spins when we compute cluster contributions in addition to the quantum evolution arising from interactions within the cluster.  In particular, Eq.~(\ref{eq:dynMapWithExtFlipFlops}) combines both effects (internal and external) using a Trotter-like approximation.  If we don't expect quantum effects to be significant, we can use an approximation that evolves clusters solely through a rate equation accounting for stochastic flip-flops. Specifically, we can replace Eq.~(\ref{eq:dynMapWithExtFlipFlops}) with
\begin{equation}
\cal{M}_{\cal{S}, \Upsilon'}'(T) = \exp\left({\bf \Gamma}_{{\cal S}, \Upsilon', \Upsilon} T\right),
\label{eq:dynMapClassical}
\end{equation}
where ${\bf \Gamma}^{\rm ext}_{{\cal S}, \Upsilon', \Upsilon}$ is the rate matrix determined from flip-flop rates with external spins \emph{and} internal spins as a superoperator given appropriate restrictions from $\Upsilon$ (to limit the pairs contributing flip-flop rates)  and $\Upsilon'$ (to limit inclusion of effective magnetic fields from external spins when calculating the flip-flop rate of a given pair).

It is advisable to calculate estimates without the classical approximation as well for comparison.  Such comparisons may be used to determine the regimes in which a classical approximation is appropriate.  This may be of interest for philosophical reasons, but also to be able to speed up calculations in regimes that have been determined to be safely classical.  Also, the classical approximation is useful when computing power spectral densities (PSDs) because the Fourier transform (FT) of autocorrelation functions based upon rate equations may be computed analytically~\cite{Gorman2012}.  We will discuss this further in Sec.~\ref{sec:experiments} where we apply this technique.

\begin{figure}
\includegraphics[width=\linewidth]{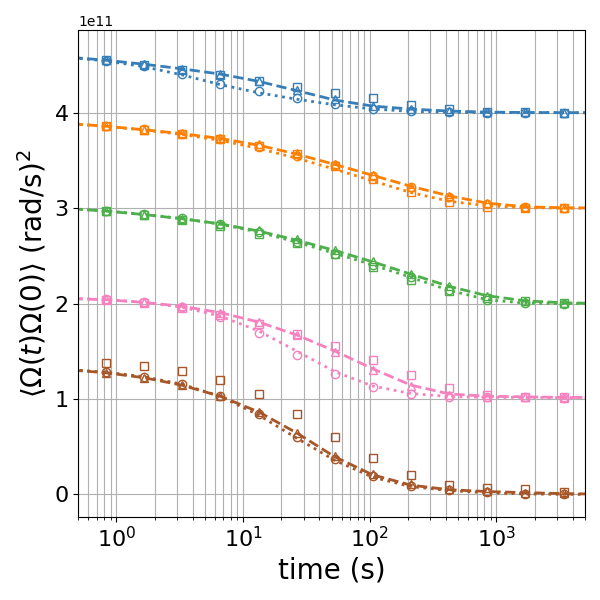}
\caption{
\label{fig:classical_comparison}
Autocorrelation functions for five different random instances of nuclear isotope locations for 500~ppm $^{29}$Si enrichment and same quantum dot shape, $t_{\rm on}$, and $t_{\rm off}$ as Figs~\ref{fig:ff_comparison} and \ref{fig:Bfield_comparison}, and
a magnetic field along the [001] crystallographic direction.
We include external flip-flops and external spin awareness with full cluster state averaging and 5 random spin states as in Fig.~\ref{fig:ff_comparison}~(b).
Offsets of $1e11~\rm{(rad/s)}^2$ separate the curves for the different cases.
Figs~\ref{fig:ff_comparison} and \ref{fig:Bfield_comparison} correspond to the blue (top) instance.  
Dashed (dotted) curves include clusters up to size 3 (2).  Classical approximations are shown with circle, triangle, and square markers for clusters up to size 2, 3, and 4 respectively.  There is good agreement between circles and dotted lines and between triangles and dashed lines demonstrating the validity of the classical approximation.}
\end{figure}

In Fig.~\ref{fig:classical_comparison} we compare results with and without the classical approximation. This approximation works very well for our examples at 500~ppm $^{29}$Si across six different random instances of isotope locations.
Deviations would become more apparent with increasing sparsity of the spin bath; with many bath spins, the effects of quantum coherence in the bath tend to wash out quickly.
This figure also demonstrates reasonable coherence with increasing cluster size from 2 to 4.  We only generated 4-cluster results under the classical approximation justified by the demonstrated validity of this approximation for smaller cluster sizes. The reasonably good agreement between 3-cluster and 4-cluster results justifies stopping at 3-clusters for most of our calculations and showing 2-cluster results to convey a pessimistic sense of accuracy. 

We observe that the classical approximation provides a significant calculation speed advantage by a factor of roughly 3, 8, and 21 respectively for clusters up to size 2, 3, and 4. 
The time to compute a 3-cluster result with external spin awareness ($\Upsilon' = \{C \in \Upsilon~{\rm s.t.}~\|C\|=2\}$) was about 13 minutes per initial spin state using 10 cpus.  For 4-cluster results we estimate it should take about 9 hours per initial spin state using 10 cpus (but less than 30 minutes under the classical approximation).

\section{Experimental validation}
\label{sec:experiments}

As a test of our newly developed methods, we compare our calculations to a diverse set of experimental results. The experimental results are from several different groups, including our own, and include both enriched and natural isotopic abundance silicon material in either silicon metal-oxide-semiconductor (SiMOS) or Si/SiGe heterostructure spin qubit devices.
We compare directly with autocorrelation functions as well as 
$T_2^*(t_{\rm avg})$ and power spectral densities that theoretically derive from autocorrelation functions. The magnetic noise data come from Ramsey experiments of either single-electron Loss-Divincenzo~(L-D) qubits~\cite{Loss1998} or singlet/triplet (S-T) qubits and employ Pauli spin blockade (PSB) read-out~\cite{Petta2005}.
The model includes the spinful isotopes of $^{29}$Si as well as $^{73}$Ge where relevant.
Unless otherwise stated, displayed simulations employ all of the techniques of Sec.~\ref{sec:method} except the classical approximation (Sec.~\ref{sec:classical}). 

\subsection{Experiment variations}

Before presenting the comparisons, we describe nuances of the experimental variations: the two electron-spin qubit models (L-D versus S-T), the role of the quantum-dot occupation schedule, and modeling the contributions from $^{73}$Ge spin dynamics. The calculations for the comparisons to experimental results use the methods presented in this paper with parametric details from the specific experiments as inputs.  Table \ref{tab:schedules} and Table~\ref{tab:parameters} consolidate the parameters for the various experiments and will be referenced throughout.
Our calculations do not rely on any fitting parameters.

\subsubsection{Models for electronic-spin qubits}
\label{sec:qubit_models}

Ramsey experiments for L-D qubits are proceed by polarizing the electron spin along an applied magnetic field, rotating it perpendicular to the applied field via electron spin resonance (ESR), allowing the spin to dephase due to magnetic noise over a period of time, rotating back via ESR, and then measuring its state via PSB.  Our model of the nuclear spin bath evolution starts with the electron perpendicular to the applied magnetic field where it can be regarded as being in an equal superposition of up and down as in Eq.~(\ref{eq:rho_0e}).

Ramsey experiments for S-T qubits are proceed by initializing the electron pair into a singlet state (the ground state when the exchange interaction is strong), then disengaging the exchange interaction to allow the spins to dephase relative to each other due to local magnetic noise over a period of time, and finally performing PSB readout.  Our model of the nuclear spin bath evolution starts with the exchange interaction essentially turned off (negligible).  We assume that each electron interacts with its own independent nuclear bath via contact HFI (long-range dipolar interactions are relatively small).  From the perspective of each independent bath, we can disregard the other electron spin.  Tracing out this extraneous degree of freedom (from the other electron spin), we have
\begin{eqnarray}
\nonumber
\rho_1 &=& \Tr_2{\left(\rho\right)} \\
\nonumber
&=&\Tr_2{\left[\left(\lvert \uparrow \downarrow \rangle + e^{i \phi'} \lvert \downarrow \uparrow \rangle \right) 
\left(\langle \uparrow \downarrow \rvert + e^{-i \phi'} \langle \downarrow \uparrow \rvert \right) /2 \right]} \\
&=& \frac{1}{2} \lvert \uparrow \rangle \langle \uparrow \rvert + \frac{1}{2} \lvert \downarrow \rangle \langle \downarrow \rvert,
\end{eqnarray}
where, for clarity, we use $\phi'$ as distinct from $\phi$ used in Eq.~(\ref{eq:rho_0e}).
This maximally mixed state of $\rho_1$ (and equivalently $\rho_2$) differs from Eq.~(\ref{eq:rho_0e}) which assumed a coherent state parameterized by $\phi$, but recall that the nuclear spin evolution was completely independent of $\phi$.  For this reason, our model of S-T qubits is the same as our model for L-D qubits except that the nuclear spin bath essentially doubles in size for the S-T qubits.  That is, we must include the dynamics and effects of nuclear spins in both of the independent nuclear spin baths.

Another experimental nuance applies to the S-T qubits. One QD has an occupation of 3 electrons in order to leverage the larger orbital splitting compared with valley splitting, used to increase the PSB readout fidelity. The 2 lower-energy electrons form a closed shell and do not interact, leaving one effective spin-1/2 electron. The only impact on our model is that the quantum dot size will be larger than would be expected if the dot was truly occupied by only one electron.

\subsubsection{The quantum-dot occupation schedule}

The quantum dot occupation schedule is a necessary input parameter for the model, consisting of either unoccupied ($t_\mathrm{off}$) or occupied ($t_\mathrm{on}$) states. During an unoccupied period, the ensemble of space-fixed, precessing nuclear spins undergoes free evolution, solely influenced by intrinsic magnetic-dipole interactions and the extrinsic Zeeman interaction. 

Free evolution is interrupted by occupied periods ($t_{\mathrm{on}}$), during which the HFI modulates the evolution of the nuclear spin bath by suppressing nuclear flip-flops. The local HFI experienced at each nucleus shifts its energy out of resonance with other nuclei comprising the bath, as well as introduces non-unitary evolution due to the random state of the occupying electron.  As a consequence, the dephasing behavior of the system is strongly influenced by the dot occupation schedule. The impact of the timing sequence can be subtle and is often neglected by assuming the nuclear bath noise can be characterized independently from the manner by which it is probed. 

Here we highlight an observable manifestation of this effect. Analogous to the well-studied phosphorus donor in Si\cite{Madzik2020}, where the presence of an electron has been shown to have a strong freezing effect on the nuclear spin bath, an electrostatically-defined QD wavefunction is expected to produce a comparable slowing of nuclear dynamics. While this effect will be weaker due to the more moderate grade of a quantum-dot wavefunction (e.g. Gaussian) as compared to a sharply-peaked donor, we nevertheless expect an observable retardation of the nuclear dynamics when the dot is occupied by an electron.
This effect will be demonstrated both experimentally and theoretically in Secs.\ref{sec:inhouse_occupation_dependence} and Sec.~\ref{sec:predictions}.

In contrast to the problem formulation of Sec.~\ref{sec:formulation}, in practice the dot occupation schedule is seldom strictly deterministic or periodic.
Dot occupations can depend upon stochastic measurement outcomes, such as in experiments that use PSB readout.
To make proper comparisons with experiments, our method accounts for a more elaborate loading schedule with stochastic occupations. Specifically, we can affect a sequence of multiple occupied-then-unoccupied durations to form a full period. As an example of this versatility, Table~\ref{tab:schedules} shows schedules pertinent to experiments we compare against in Sec.~\ref{sec:experiments}.  We provide details of the experimental schedules that inform these simulation schedules in corresponding subsections of Sec.~\ref{sec:experiments}.

In our implementation, this sequence is specified as ordered lists of durations and occupancies that together define a single period. The total duration of all loaded and unloaded intervals must equal the prescribed period, where transitions between loaded and unloaded configurations are assumed to occur instantaneously. Furthermore, for each step in the sequence, there may be multiple stochastic possibilities having the same total duration, such that the overall period is fixed regardless of the stochastic outcome. At present, the dot occupation is treated as binary corresponding to switching between two configurations in configuration space.  As an example in the singlet/triplet context, \((3,1)\) and \((4,0)\) correspond to occupied and unoccupied states, respectively.

\begin{table*}
\centering
\caption{Compound schedules pertinent to corresponding subsections used in measurements experimentally and comparable schedules used in simulation as described in the main text listed here for convenience in making comparisons.
\label{tab:schedules}}
\begin{tabular}{l c c c c c c c c c c c c c c c}
\hline\hline 
\multicolumn{2}{c}{Measurement}  &&  Probability  && \multicolumn{2}{c}{Preload ($\mu$s)} && \multicolumn{2}{c}{Control and readout ($\mu$s)} && \multicolumn{2}{c}{Idle ($\mu$s)} && \multicolumn{2}{c}{Total ($\mu$s)}  \\
\cline{1-2} \cline{6-7}\cline{9-10}\cline{12-13} \cline{15-16}

Section/id & Ref.\ &&     &&  ON &  OFF  &&  ON &  OFF  &&   ON &  OFF  && ON & OFF   \\      
\hline
\ref{sec:Foster_T2star}       &\cite{Foster2024}
                && 1.0 &&     &       &&     &      &&       &       & \vline & 93e3\footnotemark[1] & 373e3  \\ 
\hline
\ref{sec:SiGePSD}             &\cite{RojasArias2024}
                && 1.0 &&     &       &&     &      &&      &        & \vline & 30  &  13  \\
\hline
\ref{sec:UNSW_autocorrelation}/A&\cite{Stuyck2024} 
                && 0.5 &&  6  & 227.5 &&  155  &  80  &&      &    & \vline & 161 & 307.5  \\
        &       && 0.5 &&  6  & 227.5 &&  235  &   0  &&      &    & \vline & 241 & 227.5  \\
\hline
\ref{sec:UNSW_autocorrelation}/B&\cite{Stuyck2024} 
                && 0.5 &&  6  & 227.5 &&  255  &  80  &&      &    & \vline & 261 & 307.5  \\
        &       && 0.5 &&  6  & 227.5 &&  335  &   0  &&      &    & \vline & 341 & 227.5  \\
\hline
\ref{sec:inhouse_occupation_dependence}/A     & \cite{Foster2024}
                && 0.5 &&  40 &   10  &&  10.6   & 60 && 1000 & 0 & \vline &  1.05e3   &  70 \\
        &       && 0.5 &&  40 &   10  &&  70.6 &  0 && 1000 & 0  & \vline &          1.11e3  &  10 \\
\hline
\ref{sec:inhouse_occupation_dependence}/B     & \cite{Foster2024}
                && 0.5 &&  40 &   10  &&  10.6   & 60 && 0 & 1000 & \vline &  50.6   &  1.07e3 \\
        &       && 0.5 &&  40 &   10  &&  70.6 &  0 && 0 & 1000  & \vline &          111  &  1.01e3 \\
\hline\hline
\end{tabular}
\footnotetext[1]{The experiment of Ref.~\ref{sec:Foster_T2star} involves Ramsey measurement sequences, each with a duration of 93~ms, spaced by about 373~ms of computer processing time during which the quantum dots are unoccupied.  We use various proxy models as depicted in Fig.~\ref{fig:T2star_Foster_schedules} for approximating the effects from both $^{73}$Ge and $^{29}$Si.}
\end{table*}

\subsubsection{Model for the $^{73}$Ge nucleus}

Unlike the spin-1/2 $^{29}$Si nucleus, which is free of any quadrupole moment, $^{73}$Ge is spin-9/2 and therefore its quadrupole moment has a significant impact on its dynamics. The quadrupole interaction term in the Hamiltonian is $\hat{I}_{n} {\bf Q}_n \hat{I}_{n}$ for each $^{73}$Ge labeled by $n$ where ${\bf Q}_n$ is its quadrupole tensor. In the calculations reported here, we employ a phenomenological model for the $^{73}$Ge quadrupole tensor derived from fitting experimental data ~\cite{Kerckhoff2021}. Specifically, quadrupole-splitting values are drawn from a Lorentzian distribution having a width of $10$ krad/sec.  

Accounting for the structure of the quadrupole tensor reveals nontrivial dynamical contributions of individual $^{73}$Ge that interact directly with electron-spin qubits. This manifests primarily as 1-cluster contributions in our cluster calculations.  At the same time, differences in quadrupole tensors cause non-resonant internuclear interactions among $^{73}$Ge nuclei which cause 2-cluster contributions to be negligible. For this reason, we report only 1-cluster contributions from $^{73}$Ge. Meanwhile the 1-cluster contributions from $^{29}$Si are vacuous in our model. Numerical tests and comparison with experiment have confirmed that 2-cluster contributions from $^{73}$Ge may be neglected.

\begin{table*}
\centering
\caption{Model parameters of corresponding subsections reflecting experiments to the best of our knowledge and used simulations.  These include the magnetic field strength and orientation, the electronic system used to encode the qubit, the Si/SiGe well and ellipsoidal quantum-dot (QD) parameters, and alloy composition internal and external to the well. A comparison is also provided of measured ergodic T$_2^{\ast}$ times and those generated by our model.
\label{tab:parameters}
}
\begin{tabular}{c c c c c c c c c c c c c c c c c c c c c c c} 
\hline\hline
\multicolumn{2}{c}{Measurement} 
&& \multicolumn{2}{c}{B field (mT)} 
&& \multicolumn{2}{c}{qubit} 
&& \multicolumn{4}{c}{QD dimensions (nm)\footnotemark[1]} 
&& \multicolumn{2}{c}{Ge conc.\ (\%)} 
&& \multicolumn{2}{c}{Si enrich.\ (ppm)} 
&& \multicolumn{2}{c}{Ergodic $T_2^*$ ($\mu$s)} \\ 
\cline{1-2}\cline{4-5}\cline{10-13}\cline{15-16}\cline{18-19}\cline{21-22}
Section & Ref. && Str. & Orient. 
&& \multicolumn{2}{c}{type\footnotemark[2]} 
&& $w_{\mathrm z}$\footnotemark[3] & $r_x$\footnotemark[4] & $r_y$\footnotemark[4] & $\tau$\footnotemark[5]
&& Inter.\  & Exter.\  && Inter.\  & Exter.\  && Meas. \ & Model\footnotemark[6] \\
\hline
\ref{sec:Foster_T2star}     & \cite{Foster2024}      && 50 & [110] && S-T & SiGe && 5.6& 14,40 & 10,10  & 0.125 && 0 & 30  && 800 & 4.7e4 && 
3.77 & $3.8 \pm 0.2$ \\
\ref{sec:SiGePSD}\footnotemark[8] & \cite{RojasArias2024}  && 545 & [110] && L-D & SiGe && 15 & 13.9 & 2.7 & 2 && 0 & 30 && 4.7e4 & 4.7e4 && 
${\sim}1$ & $1.2$ \\
\ref{sec:UNSW_autocorrelation}  & \cite{Stuyck2024} && 700 & [110] && L-D & MOS && 0.6\footnotemark[7] & 4,10 & 4,10 & -- && -- & -- && 800 & 800 && 1.8 & $1.7\pm0.5$ \\
\ref{sec:UNSW_autocorrelation} & \cite{Steinacker2025} && 700 & [110] && L-D & MOS && 0.6\footnotemark[7] & 4,10 & 4,10 & -- && -- & -- && 400 & 400 && 4.06~\& 1.94 \footnotemark[9] & $3.0\pm1.4$ \\
\ref{sec:inhouse_occupation_dependence}    &    \cite{Foster2024}\footnotemark[10]     && 10 & [110] && S-T & SiGe && 5.6& 14,40 & 10,10  & 0.125 && 0 & 30  && 800 & 4.7e4 && 
4.09 & $3.8 \pm 0.2$ \\

\hline\hline
\end{tabular}
\footnotetext[1]{The vertical dependence of the quantum-dot (QD) wavefunction was obtained by solving the 1D Schr\"odinger equation with the potential energy proportional to the Ge concentration up to 150~meV.}
\footnotetext[2]{Classified as Loss-DiVincenzo (L-D) or singlet-triplet (S-T) qubit encodings hosted by a metal oxide semiconductor (MOS) interface or SiGe well.}
\footnotetext[3]{The MOS width is the smallest length which supports a state. The Si/SiGe well width is the distance between the interface midpoints.}
\footnotetext[4]{The lateral quantum-dot extents are denoted as $r_{x}$ and $r_{y}$, and where $r_{x}=r_{y}$ the radius is denoted in the text as $r_0$.}
\footnotetext[5]{The interface was modeled as a logistic sigmoid parameterized by $\tau$ following the convention of Ref.~\cite{Dyck2017}.}
\footnotetext[6]{Computed from HFI strengths and the spin quantum numbers of nuclear isotopes via Eq.~(10) in Ref.~\cite{Kerckhoff2021}.}
\footnotetext[7]{Value computed using the reported electric field strength of 30 MV/m}
\footnotetext[8]{Corresponding to `qubit L of D2' as labeled in Rojas-Arias et al. (Ref.~\cite{RojasArias2024}).}
\footnotetext[9]{The reported values of $T_2^*$ in Ref.~\cite{Steinacker2025} with a maximum of $40.6~\mu$s used real-time feedback drift-compensation.  The measured ergodic $T_2^*$ values were derived from $T_2^* = \sqrt{2 / \langle \Omega(0) \Omega(0) \rangle}$ [See Eq.~(\ref{eq:T2star}], where the former (latter) is from Stuyck et al. (Steinacker et al.). }
\footnotetext[10]{The same Intel-provided device as Ref.~\cite{Foster2024} but with different measurements performed at SNL corresponding to an experiment with single-shot Ramsey measurements at a fixed wait time.}
\end{table*}

\subsection{Comparing with experiments}
\label{sec:experiments}

We now compare our simulations with four sets of experiments.  The first and last are in-house (SNL) experiments on Si/SiGe spin qubit devices with enriched 800~ppm $^{29}$Si quantum wells located in between natural SiGe barriers, provided by Intel Corporation~\cite{Foster2024}. The second consists of experiments in a Si/SiGe device with natural isotopic concentrations performed at RIKEN \cite{RojasArias2024}, whereas the third are experiments on SiMOS devices enriched to 800~ppm and 400~ppm $^{29}$Si performed at UNSW~\cite{Stuyck2024, Steinacker2025}.  The first comparison demonstrates that the model accurately estimates the time required to determine $T_2^*$. The second shows compelling agreement between measured and modeled PSDs over a wide range of frequencies.
Both the $T_2^*$ and PSD quantities are directly derived from the correlation function computed by our model; for the derivations, see Appendix~\ref{appendix:T2star} and Appendix~\ref{appendix:classical_PSD}, respectively. The third set of experiments demonstrate real-time drift compensation using fast field-programmable gate array (FPGA) hardware, where we are able to compare autocorrelation functions directly for these experiments. We again compare directly to autocorrelation functions in the final set of experiments, which we designed in-house to investigate the impact of the electron spin on the noise characteristics of nuclear spin bath.

\subsubsection{Saturation of $T_2^*$ in an enriched Si/SiGe device}
\label{sec:Foster_T2star}

\begin{figure*}
\includegraphics[width=0.9\linewidth]
{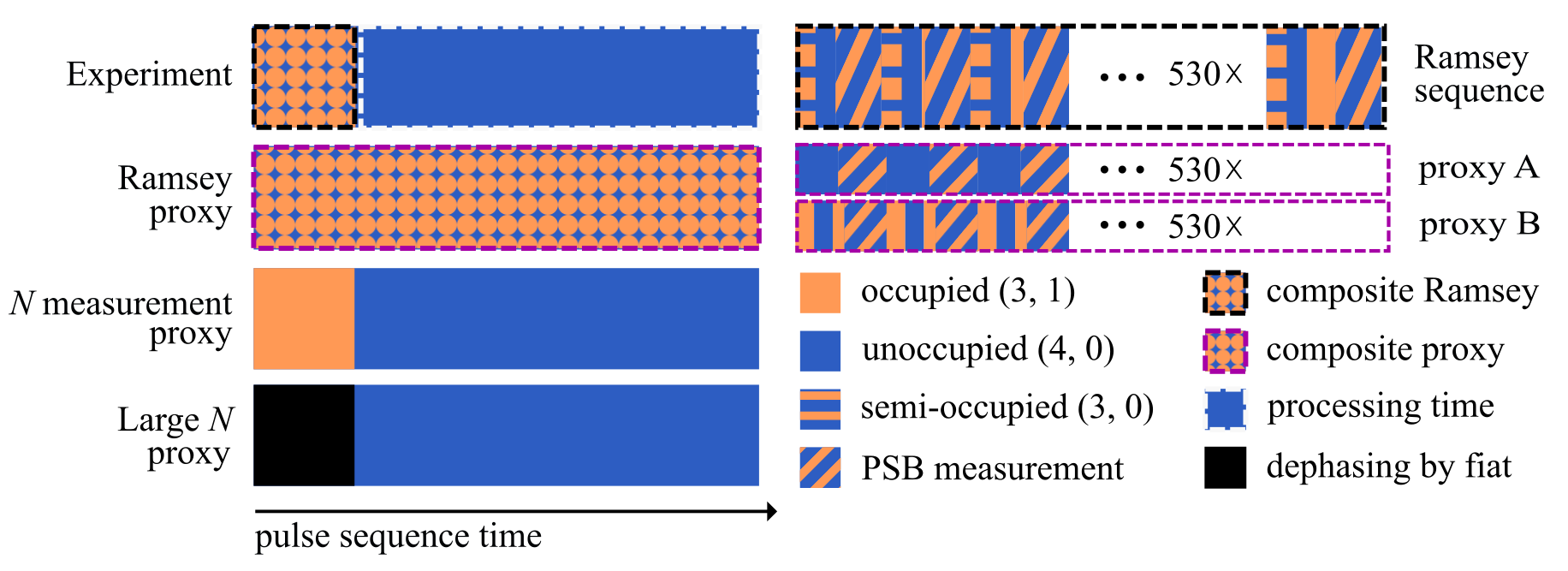}
\caption{
\label{fig:T2star_Foster_schedules}
Block schedules pertinent to Sec.~\ref{sec:Foster_T2star} for the experiment as well as simulated proxy schedules.  The experiment (first row, left) involves a sequence of Ramsey measurements with 53 wait times ranging from 10~ns to 24~$\mu$s that are each averaged over 10 shots at each wait time spanning a total duration of 93~ms (outlined in dashed black, detailed in first row, right). This sequence is repeated but interrupted by 373~ms of computer processing time during which the quantum dots are in the effectively unoccupied (4, 0) charge occupation state (long blue block).  
The Ramsey proxy schedules A (most unoccupied) and B (most occupied), are shown outlined in gray dashed (second row, detailed in two sub-rows directly right). 
$N$ repeated Ramsey measurements are modeled as refreshed electrons in between measurements, followed by computer processing time to focus on the nuclear spin
flip-flopping dynamics (of $^{29}$Si pairs)
during the long uninterrupted occupation time ($N$ measurement proxy), and
the large $N$ limit (large $N$ proxy) is approximated by ignoring the nuclear dynamics during the 93 ms of the Ramsey sequence except to set the off-diagonal components of nuclear state density matrices to zero to effect complete dephasing. The rows on the left are drawn to scale (total time 93 + $\sim$373 ms), while the rows on the right are not drawn to scale and are used for a conceptual illustration only.
}
\end{figure*}

\begin{figure}
\includegraphics[width=\linewidth]
{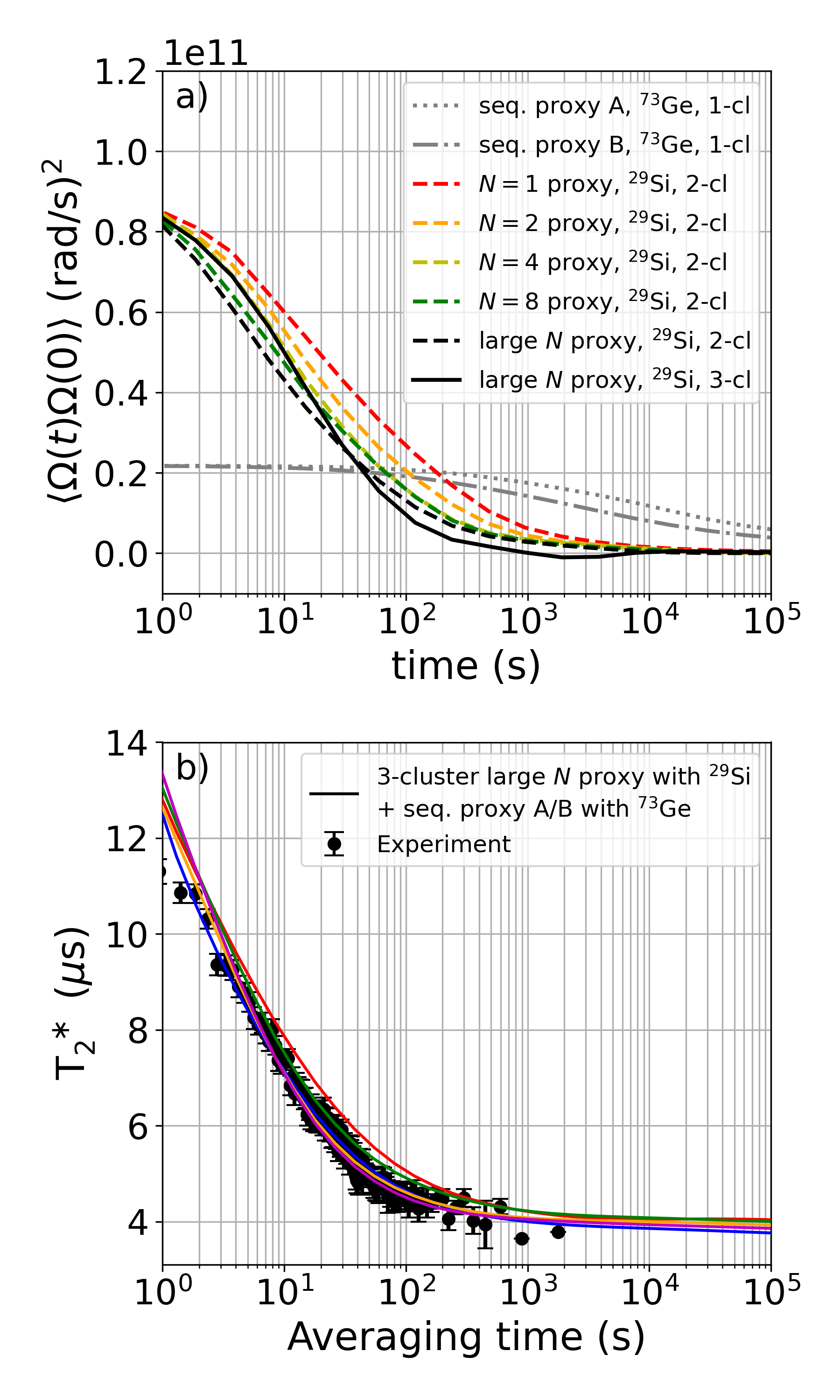}
\caption{
Comparing effective $T_2^*$ times measured as a function of data collection time from a singlet/triplet qubit [Ref.~\cite{Foster2024}] against simulation results, related to autocorrelation functions according to Appendix~\ref{appendix:T2star}.
\textbf{a)} Examines the autocorrelation functions of simulations for a particular random configuration of isotope locations and a particular initial spin state for the proxy models illustrated in Fig.~\ref{fig:T2star_Foster_schedules} using $n$-cluster ($n$-cl) approximations with $n=1,2,3$ and 2 pairs per spin (applicable beyond 1-cl).
The $^{73}$Ge is best approximated with the A/B Ramsey sequence proxies.  The $^{29}$Si contribution is significantly larger, decays more quickly, and is best approximated using the large $N$ proxy model.
\textbf{b)} $T_2^*(t_{\rm avg})$ from the experiment (black) and simulations (colors) of five random instances of isotopic configurations with a single randomly generated initial nuclear spin state. 
Error bars are standard error of the means.  Simulations result from adding contributions of 3-cluster
$^{29}$Si that use the large $N$ proxy with 2 pairs per spin and 1-cluster $^{73}$Ge using the Ramsey sequence A/B proxies (both A and B proxy results are shown but differences are negligible).  These combined autocorrelation functions are converted to $T_2^*(t_{\rm avg})$  according to Appendix~\ref{appendix:T2star}.
}
\label{fig:T2star_Foster_comparison}
\end{figure}

Experimental determination of the ergodic $T_2^*$ in quantum dot devices provides important verification of device fabrication integrity (e.g., to verify isotopic concentrations).  To ensure the ergodic value is reached, experiments require sufficient time to allow the nuclear spin bath to adequately explore its state space.  Our autocorrelation function calculations allow us to predict the time required to reach this ergodic value.  
To test our ability to make this prediction, we compare modeling and experimental calculations of averaging time-dependent $T_2^*$ whose relationship to Overhauser autocorrelation functions is described in Appendix~\ref{appendix:T2star}.  
Relevant parameters, determined for the device and used in our simulations, are shown in the corresponding entry of Table~\ref{tab:parameters}. Quantum dot dimension estimations~\cite{Neyens2024} were made as appropriate for the (3, 1)-(4, 0) charge occupations used in the experiment.

A typical $T_2^*$ Ramsey measurement involves a series of wait times that map out free induction decay of a S-T qubit in order to realize the dephasing time due to magnetic noise. The schedule used in our experiment is depicted at the top of Fig.~\ref{fig:T2star_Foster_schedules}, consisting of several dot occupations in order to prepare, evolve, and measure the S-T qubit. Following each Ramsey sequence, there is a $\sim$373 ms duration for computer processing and data averaging, during which the dots are unoccupied.  Simulating the full details of this schedule would be very difficult computationally; a full periodic cycle contains 530 measurements (53 wait times~$\times$~10 shots each) that would need to be computed separately.  To overcome this challenge, we use proxy models that are also depicted in 
Fig.~\ref{fig:T2star_Foster_schedules}.  We consider the dynamics of each nuclear species separately and then sum their autocorrelation functions together. For $^{73}$Ge, its dynamics are best modeled using proxies for the Ramsey sequence itself.  Due to the quadrupole tensors of the nuclear spins, the quantization axes are strongly dependent on the dot occupation and electron spin states.  For this reason, the $^{73}$Ge are most dynamic in time when the dot occupation is frequently changing and the electrons are regularly measured and refreshed.  We use the extreme of mostly unoccupied (occupied) for sequence proxy A (B), where we simplify the Ramsey experiment by replacing incremented wait times with identical repeated pulse sequences. We maintain the number of measurement cycles as in the experiment to faithfully preserve electron dynamics, but rescale the cycle durations such that the 530 repeated measurements span the entirety of a 93 ms Ramsey sequence plus 373 ms of computer processing time.  Confirmed by modeling (not shown), we assume the $^{73}$Ge dynamics during the 373~ms computer processing time is negligible in order to justify rescaling across this time period.  An accurate simulation of the $^{73}$Ge contribution is expected to lie in between, or in close proximity to these extremes, which are shown in gray dotted and dash-dotted lines in Fig.~\ref{fig:T2star_Foster_comparison}a).

The dynamics of the $^{29}$Si are best modeled using proxies that 
disregard fine details of the Ramsey measurements but appropriately capture the  373~ms computer processing time when the dots are unoccupied and the nuclear spins undergo free evolution.
In the limit of many repeated measurements, the $^{29}$Si dynamics during a Ramsey sequence converge to pure dephasing. We show this by considering a series of models that simulate the effect of $N$ evenly-spaced measurements (by periodically re-initializing the state of the electron pair) for $N=$1, 2, 4, 8. The colored dashed curves in Fig.~\ref{fig:T2star_Foster_comparison}a) show a reduction in the autocorrelation decay time as $N$ increases, ultimately converging on the black curves in the limit of large $N$. 
Compared to either extreme of proxy A or B for the $^{73}$Ge contribution, the $^{29}$Si contributions to the autocorrelation function are much larger 
and the decay of this contribution is much faster.
We sum the autocorrelation functions generated from the large $N$ proxy describing the $^{29}$Si, and from the mean of A and B proxies describing $^{73}$Ge in order
to model $T_2^*(t_{\rm avg})$ using the approach described in Appendix~\ref{appendix:T2star}, resulting in the colored curves shown in Fig.~\ref{fig:T2star_Foster_comparison}b). 
The range of outcomes due to different isotopic placements at this level of 800~ppm $^{29}$Si enrichment, shown as different colored errorbar curves, cover the experimental results well.  The predominant decay of $T_2^*(t_{\rm avg})$ is well-captured by our model of $^{29}$Si effects.  Our models suggest, however, that the experiment has not yet fully reached the saturation to the true ergodic $T_2^*$
that includes the $^{73}$Ge contribution, which has a more subtle and prolonged effect.

\subsubsection{PSD in Si/SiGe devices with natural isotopic concentrations}

\label{sec:SiGePSD}

\begin{figure}
\includegraphics[width=.95\linewidth,trim={1cm 4cm 1.1cm 1.75cm},clip]{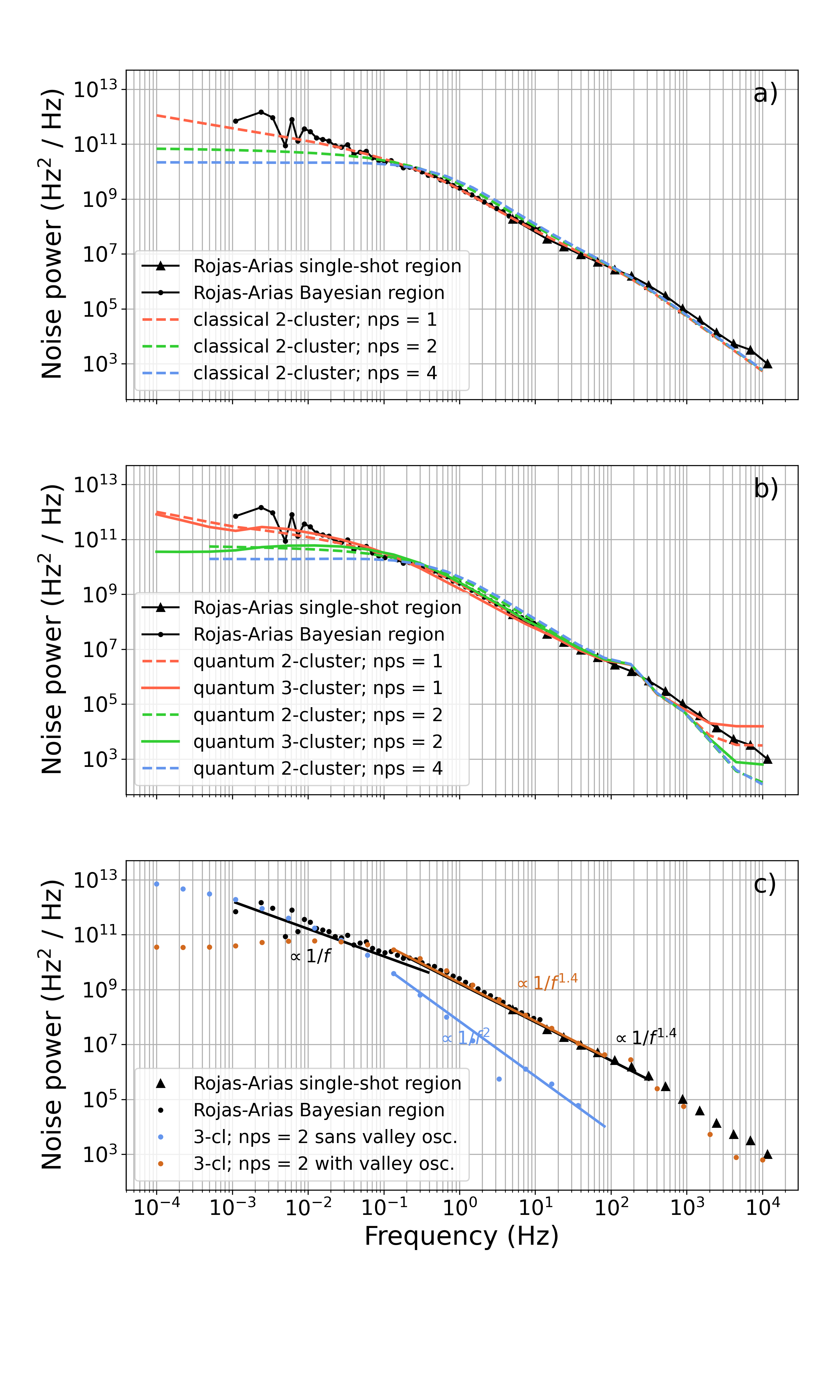}
\caption{
Power spectral densities for a single-spin qubit in an unenriched Si/SiGe heterostructure. The measured spectra (black) compare well to the present magnetic-noise simulations (colored) performed using either (a) the classical approximation or (b) the quantum CCE approximations. In (c), the importance of valley oscillations is demonstrated in the context of high-level (3-cluster) quantum simulations. The simulations were performed with both external flip-flops and external spin awareness with full cluster state averaging for one random isotopic configuration and one random default spin state (for spins external to a cluster).  Here the $N$-cluster PSD is the sum of a 2-cluster classical PSD and a quantum correction formed by taking the Fourier transform of the difference between the $N$-cluster and 2-cluster quantum autocorrelation functions.}
\label{fig:PSD_exp_comparison}
\end{figure}

We compare our simulated PSDs with the measurements of Rojas-Arias et al.\cite{RojasArias2024} in a Si/SiGe heterostructure with natural isotopic abundance, Fig. \ref{fig:PSD_exp_comparison}, using the corresponding Sect. \ref{sec:SiGePSD} schedule and model parameters listed in Table~\ref{tab:schedules} and Table~\ref{tab:parameters} respectively. We find good agreement over multiple orders of magnitude in frequency using a simple classical approximation and taking the FT analytically~\cite{Gorman2012} (see Appendix \ref{appendix:classical_PSD} for the derivation).
Corrections to the classical PSDs were formed by evaluating the FT of quantum corrections to the autocorrelation functions, exploiting linearity of the FT operation. To elaborate using an example, we can form a quantum 2-cluster PSD by adding to the classical 2-cluster PSD the FT of the difference between the quantum 2-cluster autocorrelation function and the classical 2-cluster autocorrelation function. Taking the FT of small corrections was found to be more numerically stable than computing the FT of an uncorrected quantum autocorrelation function.

Fig.\ \ref{fig:PSD_exp_comparison} is divided into three parts: a comparison between measured PSDs and purely classical simulations, quantum-corrected simulations, and the quantum-corrected simulations with and without including our model for valley oscillations. Notably, we expect quantum correlations to have a more pronounced impact for enriched systems. The environment of natural silicon is dense with nuclei, causing classical processes to dominate the dynamics. 

Good agreement is observed between the experiment and the classical PSDs throughout the frequency range spanning 0.1 to 10,000 Hz as can be seen in Fig. \ref{fig:PSD_exp_comparison}(a). In the low-frequency regime we note that the classical noise power deviates from measured values likely due to known convergence challenges of our method with a high density of nuclear spins in the natural silicon spin bath. We validate the origins of this deviation by increasing the number of pairs per spin (nps) and observing increasingly stronger deviations from the experimental data.

In Fig. \ref{fig:PSD_exp_comparison}(b), quantum corrections of both 2-cluster and 3-cluster type are introduced to correct for shortcomings of the classical approximation. We again find good agreement for the frequency range spanning 0.1 to $10^3$ Hz, indicating this regime is well converged. However, the quantum correction simulations of Fig.\ \ref{fig:PSD_exp_comparison}(b) reveal convergence challenges in the high ($>10^3$ Hz) and low ($<10^{-1}$ Hz) frequency extremes and competing effects that fortuitously cancel to make the roughest approximations the most accurate.  Increasing the nps pushes the noise power down at the frequency extremes and slightly up in the middle while increasing the cluster size which counters this for a PSD-straightening effect. The accuracy of the classical approximation is expected to fall off when moving away from natural silicon to more sparse spin-bath systems though still works fairly well at 500 ppm $^{29}$Si, as observed in Fig.~\ref{fig:classical_comparison}.

A subtle bump manifests in all quantum-corrected PSDs at around 180 Hz.  The nearest neighbor interaction is $\left( \hat{I}_m^{+} \hat{I}_n^{-} + \hat{I}_m^{-} \hat{I}_n^{+} - 4 \hat{I}_m^{z} \hat{I}_n^{z} \right) \times 91.3~\rm{Hz} \times (2 \pi \hbar)$ in the secular approximation.  We can consider the $\{|\Uparrow \Downarrow \rangle, |\Downarrow \Uparrow \rangle \}$ states of such a pair of $^{29}$Si nuclei as a $1/2-$pseudospin~\cite{Yao2006} with a maximum rotation frequency of $182~\rm{Hz}$, matching this PSD hump.   This bump is not clearly noticeable in the experiment, possibly due to non-idealities such as an imperfect magnetic field alignment relative to the lattice orientation.

In Fig. \ref{fig:PSD_exp_comparison}(c), 3-cluster quantum-corrected simulations were performed with and without valley oscillations, whereas they were included in all other simulations presented here. In their study, Rojas-Arias et al. identified the origins of the intermediate $1/f^{1.4}$ regime (i.e., 0.1-100 Hz), which bridges the traditional $1/f$ and $1/f^{2}$ regimes, as being attributable to valley oscillations in silicon.  We validate this claim by showing that the inclusion of valley oscillations in our model matches the experimental $1/f^{1.4}$ PSD observed at intermediate frequencies. Further, with valley oscillations toggled off, the  simulated intermediate regime exhibits a clear $1/f^{2}$ decay. This result reinforces the importance of valley oscillations in simulations of magnetic-noise dynamics in silicon.

\subsubsection{Autocorrelation functions of enriched SiMOS devices using drift-compensation}
\label{sec:UNSW_autocorrelation}

\begin{figure}
\includegraphics[width=\linewidth]{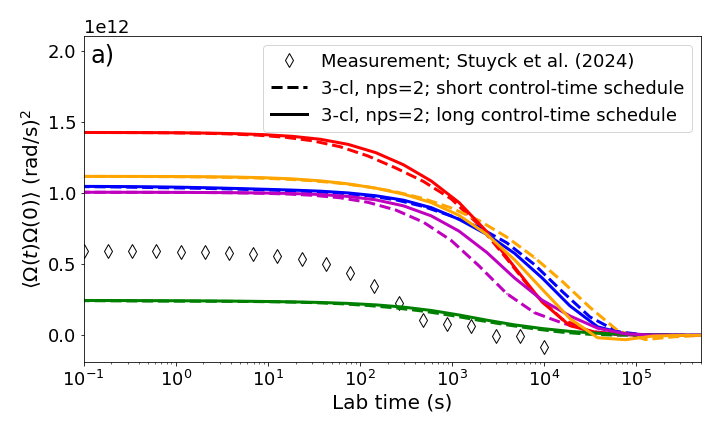}
\includegraphics[width=\linewidth]{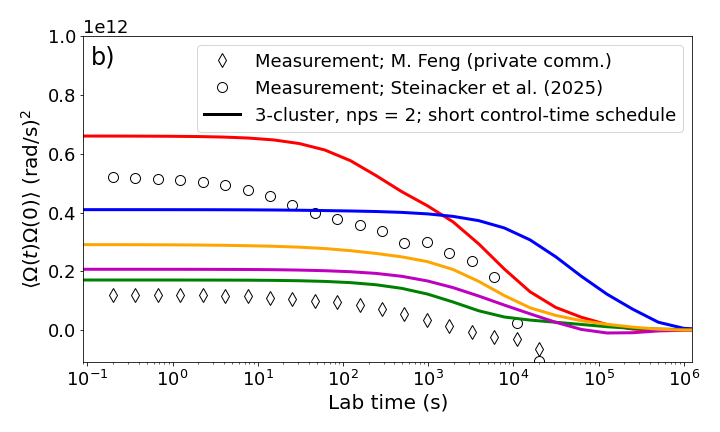}
\caption{Autocorrelation functions derived from measurements at UNSW having $^{29}$Si enrichment levels of (a) 800 and (b) 400 ppm, where the experimental data was taken from Refs\ \cite{Stuyck2024} and \cite{Steinacker2025}, respectively. Simulations were performed using the device parameters listed in Table \ref{tab:parameters} with the quantum 3-cluster CCE, 2 pairs per spin, and including magnetic influencers. Each colored line represents a unique distribution of nuclear spins on the lattice. 
Short and long control-time schedules, as described in Table \ref{tab:schedules}, were provided by UNSW. The long-time schedule adds 99.8 $\mu$s to the control step. Error bars on the measured data, computed as standard errors of the mean, were imperceptibly small and therefore omitted. 
\label{fig:T2star_UNSW_comparison}
}
\end{figure}

We now turn to drift-compensation experiments, which provides an opportunity to verify our drift-compensation predictions, which we discuss in greater detail in Sect.\ \ref{sec:predictions}. At UNSW, the performance of L--D qubits in SiMOS has been studied at two enrichment levels: 800 ppm \(^{29}\)Si in Ref.~\cite{Stuyck2024} and 400 ppm \(^{29}\)Si in Ref.~\cite{Steinacker2025}. These experiments also extend our analysis into a substantially different device regime: unlike the other cases considered here, they involve SiMOS (rather than Si/SiGe) quantum dots and higher $^{29}$Si enrichment, which means fewer spinful nuclei experience significant hyperfine interactions within the more localized electron densities. Agreement in this setting therefore illustrates that our method remains effective across a diverse range of device characteristics.

Each experiment contains two electron spin qubits on two quantum dots (one quantum dot also contains a filled shell of two additional electrons), one being addressed via ESR and the other as a reference for PSB parity measurements.  The Larmor frequency of the addressed qubit is tracked using fast FPGA hardware by attempting to rotate the addressed qubit by $\pi/2$ so that two parity outcomes should be equality probability.  Observing a prevalence of one outcome over the other indicates that the Larmor frequency has drifted and must be compensated.

Figs.~\ref{fig:T2star_UNSW_comparison}(a) and \ref{fig:T2star_UNSW_comparison}(b) show autocorrelation functions computed from temporal feedback frequencies 
for L-D SiMOS qubits fabricated by Diraq on 800 and 400 ppm $^{29}$Si wafers, respectively. The autocorrelation functions were obtained from the measured feedback-frequency difference between the two qubits by subtracting its mean, interpolating the resulting time series, and then computing the time-averaged product $\langle \delta\omega(t')\,\delta\omega(t'+t)\rangle$ over all pairs of points separated by a lag time $t$.
By taking differences of the qubit frequencies, we remove effects of external magnetic noise that are correlated between the two dots (these correlations can be substantial as seen by comparing Figs.\ 2c and 2d in Ref.~\cite{Stuyck2024}), while retaining the local effects of the respective nuclear spin baths.
The experimentally derived autocorrelation functions are compared with our simulations for five distinct random $^{29}$Si lattice configurations. The occupation schedules and measurement parameters used in the simulations are listed in Tables~\ref{tab:schedules} and \ref{tab:parameters}, respectively.

The simulated results vary substantially across different $^{29}$Si configurations, reflecting both the high enrichment and the relatively small size of the SiMOS quantum dots.  In addition, other sources of noise may be hastening the autocorrelation decay.  With these caveats, our modeling is reasonably consistent with these experiments, in terms of both the zero-time values of the correlation functions and the lab time in which the decay is predicted to occur, particularly when compared to the more recent measurements provided by UNSW in Fig.~\ref{fig:T2star_UNSW_comparison}(b). For these highly-enriched systems, a relatively large variance in the zero-time correlation function was observed; we were, however, able to bound almost all of the experimental data by running only five randomly-chosen nuclear distributions.

The experimental sequence involves two different readout schedules that differ by 100~$\mu$s during the ``control'' phase of the schedule.
Instead of simulating the sequence precisely, we performed simulations at the two extremes, with and without adding the extra 100~$\mu$s during each readout. 
These are identified as the A and B schedules for Sec.~\ref{sec:UNSW_autocorrelation} in Table~\ref{tab:schedules}.
Results of both extremes are shown in the top panel of Fig.~\ref{fig:T2star_UNSW_comparison} (dashed versus solid curves) and the difference is relatively minor, justifying 
usage of the simplified schedule for these comparisons
The schedules of Table.~\ref{tab:schedules} capture a starting occupation in (3, 1), initialization and preload in (4, 0), followed by control, load, and a reference control point in (3, 1), and finally a PSB readout which we take to be in either occupation state with equal probability.  An ESR pulse (involving either a single-frequency or an adiabatic frequency sweep in correspondence with the two readout schedules) is applied during the control phase.  Although our simulations neglect the effect of the ESR pulse on the bath, we observe a dephasing effect incurred from unknown electron spin polarizations during the control phase and assume this is sufficient to capture the electrons'  impact on the nuclear spin bath dynamics
in these experiments~\footnote{The hyperfine interaction with $^{29}$Si in these experiments should be small relative to their Zeeman energy.  Thus, the impact of the electron should be predominantly a dephasing effect rather than inducing nuclear flips}.

The $T_2^*$ values reported in Ref.~{\cite{Steinacker2025}} are in the 20-40~$\mu$s range which is much longer than the $\sim$3~$\mu$s ergodic $T_2^*$ that derives from $T_2^* = \sqrt{2 / \langle \Omega(0) \Omega(0) \rangle}$ [Eq.~\ref{eq:T2star}] given their measured feedback-frequency differences.  Time between feedback measurements in that experiment, based on data provided to us by UNSW, was typically about 25~ms with occasional excursions closer to 1~s. Utilizing Eq.\ \ref{eq:drift_compensated_T2star_def}, our processing of the experimental data returned $T_2^*=20.4\ \mu$s at a lab time of 25~ms and 7.8 $\mu$s at a lab time of 1 sec; while the former is consistent with the reported range, the value decreases substantially before reaching the ergodic limit. In Sec.~\ref{sec:predictions}, we make predictions of drift-compensated $T_2^*$, which derives directly from autocorrelation functions, for various device and schedule parameters as a function of Overhauser calibration time intervals, $\Delta t$.  Those results are consistent in predicting significant drift-compensated $T_2^*$ enhancement when $\Delta t \leq 1$~s.

\subsubsection{Context dependent noise due to dot occupation in an enriched Si/SiGe device}

\label{sec:inhouse_occupation_dependence}

In our final comparison, we revisit in-house experiments using the Si/SiGe device provided by Intel Corporation.  Schedules and parameters are again given in Tables~\ref{tab:schedules} and~\ref{tab:parameters} identified by~\ref{sec:inhouse_occupation_dependence}. 
We modify our Ramsey measurement approach here for a clear demonstration of the context dependence of the nuclear spin bath with respect to the dot occupation schedule, a less dramatic version of the nuclear spin bath freezing observed in the context of phosphorus donors~\cite{Madzik2020}.  

We apply the single-shot approach using a fixed Ramsey wait time, $\tau$, as described in Ref.~\cite{RojasArias2024} and originally proposed in Ref.~\cite{Fink2013} which offers more precise control over quantum dot occupation durations and eliminates variable computer processing time occurring in between repeated Ramsey traces during the full experiment. Here, the data transfer for computer processing occurs only after each 60,000 shot experiment, lasting about 94 seconds, is completed. These experiments are repeated 50 times over the course of about 3 hours to improve statistical certainty but the time in between repetitions does not impact the nuclear spin evolution being probed.  This approach also allows us to calculate correlation functions from the experimental measurements given $\tau \gg T_2^*/\sqrt{2}$ as described in Ref.~\cite{RojasArias2024}.  In these experiments, $T_2^* \approx$~4.0 $\mu$s and we chose $\tau=10.6~\mu$s.

The essence of the approach described in Ref.~\cite{RojasArias2024} is to evaluate the time-correlator of the single shot measurements,
\begin{equation}
C_P(t) \equiv \langle P(t_0) P(t_0 + t)\rangle - \langle P \rangle^2,
\end{equation}
where $P(t)$ is the expectation value of the single-shot measurement at time $t$.  Given $\tau \gg T_2^*/\sqrt{2}$,
\begin{equation}
C_P(t) \approx \frac{A^2}{2} e^{-4 \pi^2 \tau^2 \left( 
\langle \delta \nu^2 \rangle - \langle \delta \nu(t') \delta \nu(t' +t)\rangle \right)},
\end{equation}
where $A$ accounts for SPAM error and $\delta \nu(t)$ is the qubit-energy fluctuation at time $t$ in $2 \pi \hbar$ units~\cite{RojasArias2024}. In particular, $2 \pi \delta \nu(t) = \Omega(t)$ holds if the qubit-energy fluctuation is due to Overhauser noise.  Under this assumption, the autocorrelation function is
\begin{equation}
\langle \Omega(t) \Omega(0) \rangle \approx \langle \Omega(0) \Omega(0) \rangle + \frac{1}{\tau^2} \left(\log \frac{C_p(t)}{C_p(0)} \right).
\end{equation}
This method directly probes noise at intermediate frequencies~\cite{RojasArias2024} and provides an advantage for our purpose of using a periodic dot-occupation schedule consistent with our model assumptions.  One weakness, however, is a large uncertainty near the tail of the autocorrelation decay where $C_p(t) \ll C_p(0)$ and a small uncertainty of $C_p(t)$ can cause a large uncertainty of $\langle \Omega(t) \Omega(0) \rangle$ because of the logarithm.

\begin{figure}
\includegraphics[width=\linewidth]{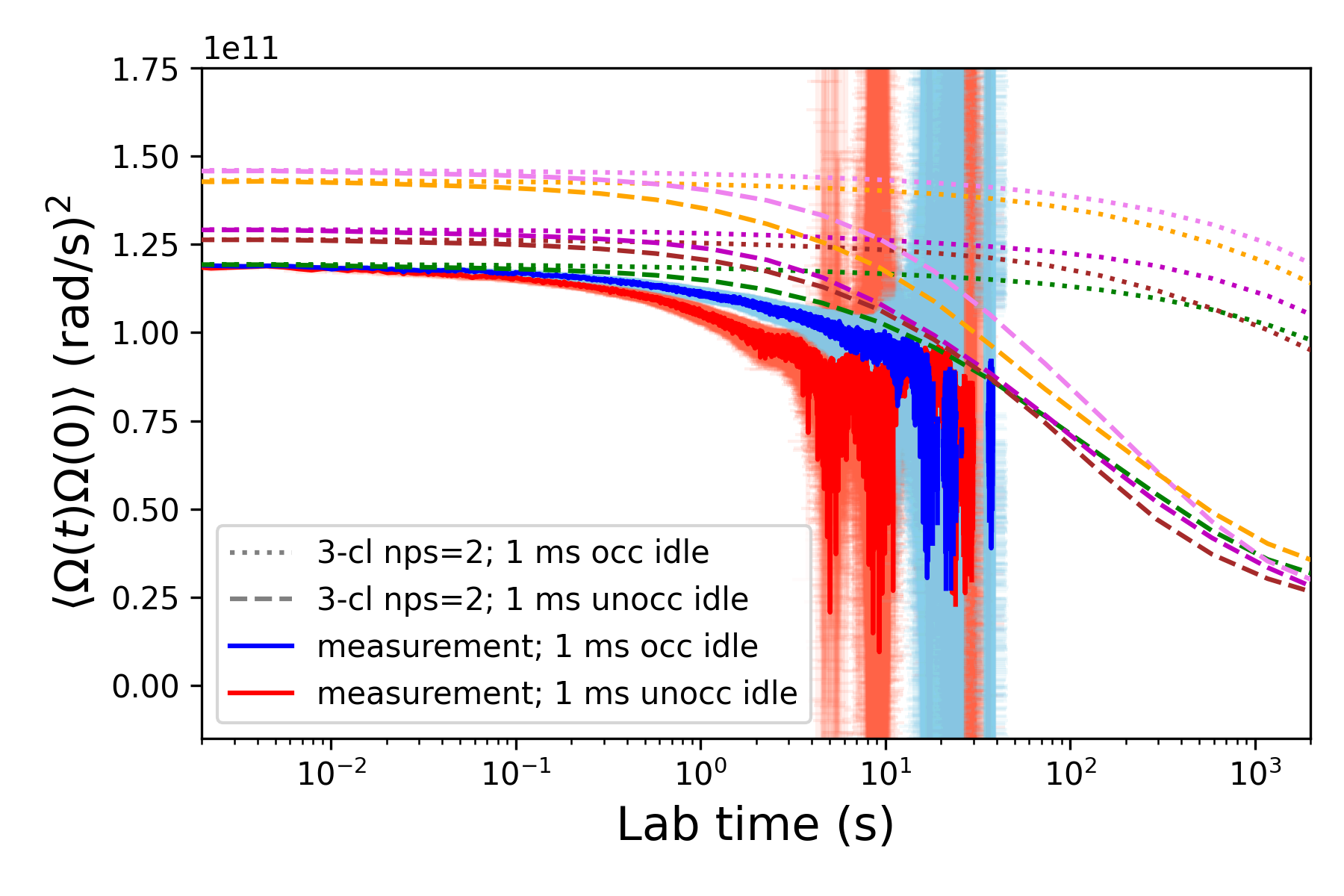}
\caption{
\label{fig:occupation_dependent_noise}
Overhauser autocorrelation functions from simulation and experiment for double-dot S/T Ramsey experiments with parameters as listed under \ref{sec:inhouse_occupation_dependence} in
Tables~\ref{tab:schedules} and~\ref{tab:parameters} for occupied (A) and unoccupied (B) idle.
Each colored line represents a unique distribution of nuclear spins on the lattice. 
The difference between these is an indication of the back-action effect of the electron on the dynamics of the nuclear spin bath.  Simulations were executed for five different instances of isotopic location of nuclear spins and averaged over five different initial spin states.
}
\end{figure}

The A and B schedules, as shown in Table~\ref{tab:schedules}, correspond to fixed-$\tau$ Ramsey measurement interspersed with idle of $1$~ms in either an occupied (3, 1) charge configuration (A) or unoccupied (4,0) charge configuration (B).  By comparing the effects of these two schedules (occupied or unoccupied idle) we can see the the context dependence of nuclear spin noise due to dot occupation.  As shown in Fig.~\ref{fig:occupation_dependent_noise}, 
this effect is significant for both experiment and theory.
The dynamics of the nuclear spin bath is substantially impacted by the electron spins in the quantum dots.
For both theory and experiment, the correlation function decay is prolonged for the occupied idle case versus the unoccupied idle case.  This is expected because occupying the dots with unpaired electrons is known to suppress the dynamics of the nuclear spin bath.  The experimental decay is significantly faster than the corresponding decay in the theoretical results suggesting that there is a noise effect in the device that is not being modeled adequately. Better agreement between theory and experiment was found for the same device when using a different schedule in Sec.~\ref{sec:Foster_T2star} (see Table~\ref{tab:schedules}) which was within a regime with a faster autocorrelation function decay; in fact, the general trend is for the model to deviate further from the experiment as decay times are prolonged, consistent with an unaccounted noise source that is setting an upper limit on the autocorrelation decay time.  This will be important to investigate in future experiments.

\section{Predicting benefits of drift compensation}
\label{sec:predictions}

Nuclear spin dynamics can be very slow in a silicon material with some enrichment.  As such, the Overhauser rotation of an electron in a quantum dot is relatively stable at short time scales but experiences long term drift that can be harmful to qubit operations. To counteract the effects of long-term Overhauser drift, we consider a drift compensation scheme. Periodic measurements of the Overhauser rotation will quantify the drift and could be used to compensate quantum gate operations (i.e., the quantum circuit).  Doing so potentially reduces enrichment demands for a given fidelity target.  But how often must the Overhauser rotation be measured for a desired fidelity as a function of enrichment?  The quantitative answer to this question is directly related to the autocorrelation of $\Omega$ that we are now able to compute given our new methods.  Regularly measuring and compensating for the Overhauser rotation changes the effective $T_2^*$ to the drift-compensated $\tilde{T}_2^*(\Delta t)$ [Eq.~\ref{eq:drift_compensated_T2star_def}] where $\Delta t$ is the time since the Overhauser was last measured 
[see Appendix~\ref{appendix:T2star} for details].  If we measure and compensate at a period of $T_M$, the effect $T_2^*$ will be bounded by $\tilde{T}_2^*(T_M)$, higher than the ergodic $T_2^*$ if $T_M$ is sufficiently short.

\subsection{Predictions}

\begin{figure}
\includegraphics[width=\linewidth]{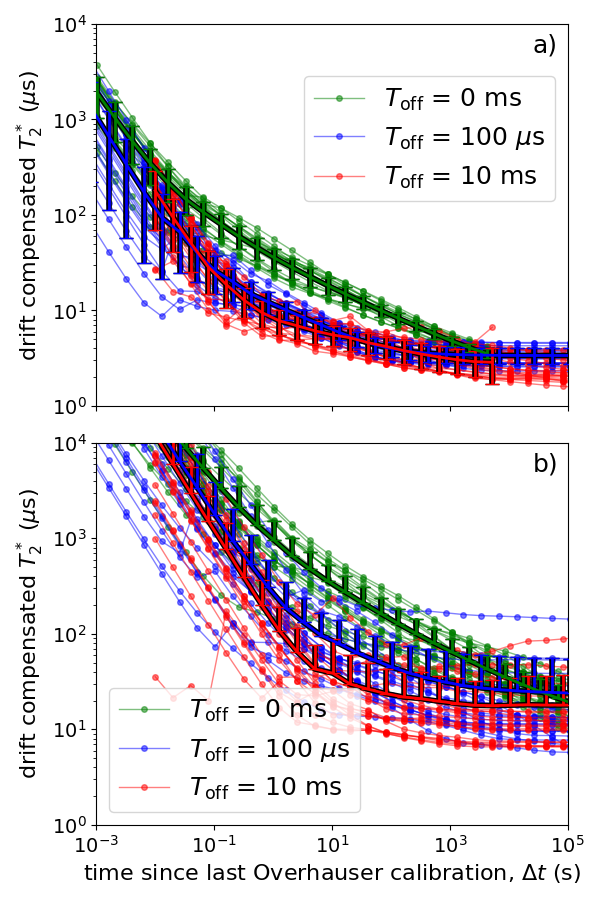}
\caption{
Simulations were performed with a silicon enrichment of (a) 500ppm and (b) 50 ppm, 
the quantum dot dimensions were $r_0 = 10$~nm and $z_0 = 5$~nm, 
and the periodic dot occupation schedule was 
$T_{\rm on} = 1\mu$s occupied and $T_{\rm off} \in \{0, 100\mu{\rm s}, 10{\rm ms}\}$ unoccupied. 
Each semi-transparent data set corresponds to a different random isotopic constellation.
}
\label{fig:Prediction1usTon}

\end{figure}

\begin{figure}
\includegraphics[width=\linewidth]{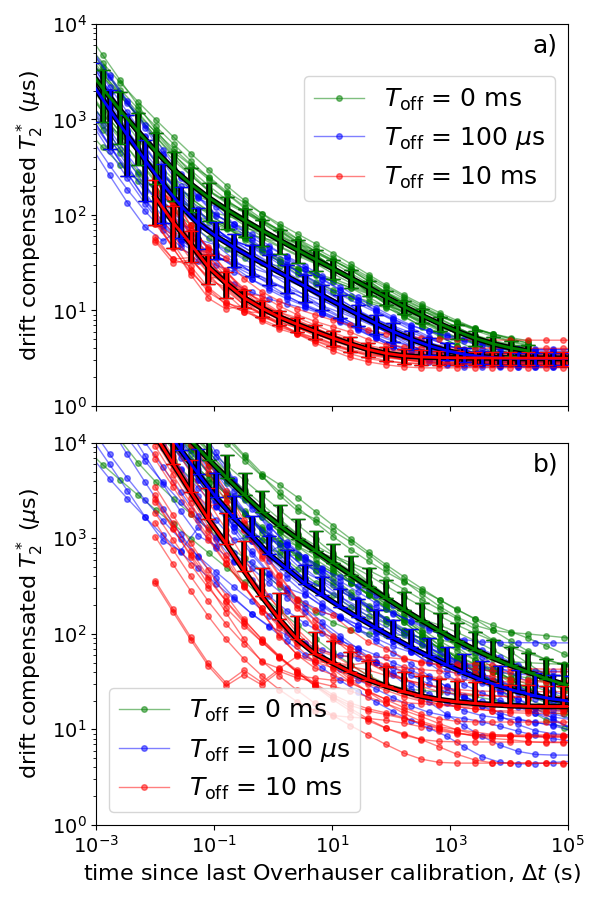}
\caption{
\label{fig:Prediction5usTon}
Simulations were performed with a silicon enrichment of (a) 500ppm and (b) 50 ppm, 
the quantum dot dimensions were $r_0 = 10$~nm and $z_0 = 5$~nm [Eq.~(\ref{eq:wavefunction})], 
and the periodic dot occupation schedule was 
$T_{\rm on} = 5\mu$s occupied and $T_{\rm off} \in \{0, 100\mu{\rm s}, 10{\rm ms}\}$ unoccupied.
Each semi-transparent data set corresponds to a different random isotopic constellation.
}
\end{figure}

\begin{figure}
\includegraphics[width=\linewidth]{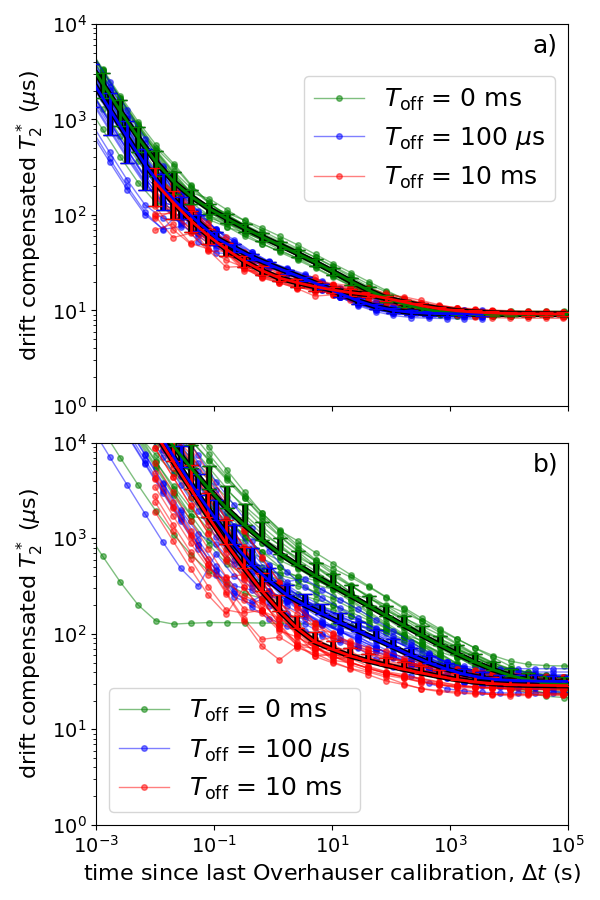}
\caption{
\label{fig:Prediction5usTon_BigDot}
Simulations were performed with a silicon enrichment of (a) 500ppm and (b) 50 ppm, 
the quantum dot dimensions were $r_0 = 20$~nm and $z_0 = 10$~nm [Eq.~(\ref{eq:wavefunction})], 
and the periodic dot occupation schedule was 
$T_{\rm on} = 5\mu$s occupied and $T_{\rm off} \in \{0, 100\mu{\rm s}, 10{\rm ms}\}$ unoccupied.
Each semi-transparent data set corresponds to a different random isotopic constellation.
}
\end{figure}

How does the drift-compensated $\tilde{T}_2^*(\Delta t)$ depend upon enrichment, duration of time an electron is on/off the dot, and the dot size?  We explore these questions (computing results for various parameter values) in Figs.~\ref{fig:Prediction1usTon}--\ref{fig:Prediction5usTon_BigDot}.  
Results for a smaller dot are shown in
Fig.~\ref{fig:Prediction1usTon} and Fig.~\ref{fig:Prediction5usTon} with
 $T_{\rm on}=1~\mu$s and $T_{\rm on}=~5\mu$s respectively.
 Results for a bigger dot with $T_{\rm on}=5~\mu$s are shown in 
Fig.~\ref{fig:Prediction5usTon_BigDot}.
Our simulations assume an ideal setting in which: (1) each electron arrives randomly polarized (e.g., half of a S-T pair) and (2) each period of alternating free and occupied durations is temporally identical and it repeats indefinitely.
These assumptions are likely reasonable in the context of quantum error correction~\cite{Nielsen2010} where syndrome measurements must be repeated with regularity. 
These simulations were performed using cluster state averaging, external spin awareness, and external flip-flops 
but did not invoke the classical approximation.
Curves in the background are 20 individual $\tilde{T}_2^{\ast}$ curves with each instance being a unique isotopic spatial realization with one default spin state each [one value of $k$ for Eq.~(\ref{eq:state_avging_with_esa})]; the foreground shows the average and standard deviation of these realizations. 

  As can be seen Figs.~\ref{fig:Prediction1usTon} -~\ref{fig:Prediction5usTon_BigDot}, minimizing $T_{\rm off}$ gives the highest values of $\tilde{T}_2^*$, but the range of outcomes is limited over the range of its extremes (zero to infinity, effectively represented by our  $T_{\rm off} = 10$ms results). The differences between $T_{\rm on} = 1\mu$s versus $T_{\rm on} = 5\mu$s are subtle on logarithmic scales when comparing Fig.~\ref{fig:Prediction1usTon} and Fig.~\ref{fig:Prediction5usTon}. The size of the quantum dot also has a subtle effect when comparing Fig.~\ref{fig:Prediction5usTon} and Fig.~\ref{fig:Prediction5usTon_BigDot}. Increasing enrichment (lowering the density of $^{29}$Si) increases $\tilde{T}_2^*$ as expected, and increases the variation for different instances of quantum dots (represented by the error bars). Our results indicate a potential significant enhancement of the effective $T_2^*$ by compensating for the Overhauser drift
  with $T_M<1$ second.

\subsection{Sn qubit}
One particular motivation for our investigation of how re-characterization of the Overhauser field and compensation can enhance $\tilde{T}_2^*(t)$ is the prospect of relaxing the isotopic-enrichment requirement within the context of our proposal for a tin nuclear-spin qubit~\cite{Witzel2022}, or any similar nuclear spin qubit system. In our proposal, spin-1/2 nuclei in silicon, specifically $^{119}$Sn or $^{117}$Sn, would be leveraged as long-lived qubits with entanglement generated by coherently shuttling electrons~\cite{Struck2024, Volmer2024} between the tin qubits using electrodes that define quantum dots.  The entangling gate is induced simply by placing an electron where it can interact strongly with a tin qubit through the HFI for a duration dictated by the hyperfine strength.  During this time, the electron decoheres through interactions with extraneous nuclear spins, inducing a dephasing error probability of $\frac{1}{2} \left( 1 - \exp{\left(-\left(T/\tilde{T}_2^* \right)^2\right)} \right)$, where $T$ is the gate time.  Typical entangling gate times expected for tin qubits are $1-5$~$\mu$s~\cite{Witzel2022}. To ensure error probabilities below 1E-3, for example, $T_2^*$ should exceed 100~$\mu$s which is likely to be realized only with devices enriched well below 50~ppm of $^{29}$Si~[See Fig.~6 of Ref.~\cite{Witzel2022}].  

For a concrete scenario that serves as motivation for this work, we consider operating multiple $^{119}$Sn (or $^{117}$Sn) qubits in silicon~\cite{Witzel2022}.  Entanglement between the tin qubits is generated by shuttling electrons from one to another.
Each tin qubit will be surrounded by a bath of extraneous $^{29}$Si spins at a concentration determined by the degree of isotopic enrichment.  While an electron interacts with any given tin qubit, it also interacts with its nearby $^{29}$Si bath, which can induce an unknown rotation on the electron spin qubit.
Here we regard the nuclear spin baths of distinct tin qubits as independent of each other, assuming the tin qubits are sufficiently far apart.  
We assume that the same electron will interact with a given $^{119}$Sn qubit only once, playing a short-term role in mediating entanglement between nuclear spin qubits. Consequently, we can trace out the electron degrees of freedom in the density matrix of the quantum state after it interacts with the nuclei. Let us further assume that the electron will be in an equal superposition of spin-up and spin-down states to maximize its efficiency in generating entanglement (since the $^{119}$Sn qubit state only impacts the electron's phase through the $\hat{S}_z \hat{I}_{nz}$ interaction).
Further, assume an error correction context in which electrons are regularly prepared, entangled with nuclear spin qubits, and then measured.  This precisely justifies our model with periodic $T_{\rm off}$ and $T_{\rm on}$ as presented in Sec.~\ref{sec:formulation}.  Thus, our $T_2^*(t_{\rm avg})$ predictions of Figs.\ 
\ref{fig:Prediction1usTon}, \ref{fig:Prediction5usTon}, and \ref{fig:Prediction5usTon_BigDot} are directly applicable to such a quantum error correction scenario for our tin qubit proposal. 

As an example scenario, assume a tin qubit (e.g., $^{119}$Sn) in a quantum dot with dimensions of $r_0 = 10$~nm and $z_0 = 5$~nm at 50~ppm $^{29}$Si 
with a HFI strength of $100~$kHz
and
assume a syndrome extraction cycle time of $100~\mu$s or less such that $T_{\rm off} \leq 100 \mu$s.  
The probability of an electron $Z$ flip during an electron-nuclear CPhase operation may be expressed as 
$\frac{1}{2} \left(1 - \exp\left(- \left( T/\tilde{T}_2^*\right)^2 \right)\right)$
where $T = T_{\rm on} = 5~\mu$s is the time of the gate operation~\cite{Witzel2022}.  The ergodic $T_2^*$ will range from 10 to 50 $\mu$s for most quantum dots of the assumed dimensions~\cite{Witzel2022}.  Without drift compensation, the electron Z flip probability would then range between $0.005$ and $0.1$.
If we periodically measure and characterize the Overhauser rotation at an interval of once every 100~ms, the 
effective $T_2^*$ with drift compensation just prior to each next re-characterization
will range between about 100~$\mu$s and 10~ms (see the blue curves in the bottom panel of Fig.~\ref{fig:Prediction1usTon} at $\Delta t = 0.1 s$).  This improvement over an order of magnitude in the effective $T_2^* = \tilde{T}_2^*(0.1 s)$ would reduce the electron Z flip probability to a range of about $1.2~\times 10^{-7}$ to $1.2~\times 10^{-3}$, a dramatic improvement.

\section{Conclusion}
\label{sec:conclusion}
 
We developed a modified version of CCE with better convergence for characterizing the long term behavior of a nuclear spin bath that periodically interacts with an unpolarized electron spin.
With this new method, we found agreement with experiments and made predictions to determine effectiveness of drift compensation of the magnetic noise generated by $^{29}$Si in quantum dots to enhance the effective $T_2^*$.
Frequently occupying a quantum dot with an unpolarized electron and keeping the unoccupied duration short slows the nuclear spin dynamics for better drift compensation (an effect previously noted for phosphorous donors in Ref. \cite{Madzik2020}).

In the context of using tin qubits in silicon as proposed in Ref.~\cite{Witzel2022}, costly enrichment requirements may be relaxed by employing drift compensation.  According to our predictions, as an example, $T_2^*$ can effectively increase by over an order of magnitude in 50~ppm $^{29}$Si if the Overhauser rotation is recharacterized every $100$~ms for a quantum dot with dimensions of $r_0 = 10$~nm and $z_0 = 5$~nm and a hyperfine strength of $100~$kHz.  This would result in a dramatic decrease of the electron Z flip probability during an electro-nuclear controlled-Z gate operation (an entangling operation between an electron spin and a tin qubit) by two orders of magnitude or more.

\section{Acknowledgements}
We acknowledge useful suggestions and conversations with Kevin Young, Noah (Toby) Jacobson, Ezra Bussmann, Thaddeus Ladd, and James Larsen.  We acknowledge Intel Corporation for supplying devices used in measurements presented in Fig.~\ref{fig:T2star_Foster_comparison} and Fig.~\ref{fig:occupation_dependent_noise} and for helping us to determine appropriate parameters to use in our models of their devices, particularly Fahd Mohiyaddin, Matthew Curry, and Nathan Bishop.  We acknowledge Juan Rojas-Arias for providing details of the experiments in Ref.~\cite{RojasArias2024} and for suggesting that we include the insightful comparison of the PSD with and without valley oscillations. 
We are grateful to MengKe Feng, Henry Yang, Santiago Ramirez, Nard Stuyck, and Paul Steinacker for providing valuable details with clear explanations about experiments performed at UNSW.  We further appreciate additional experimental data provided by MengKe and Santiago enabling more careful autocorrelation-function comparison.
Research was sponsored by the Army Research Office and was accomplished under Cooperative Agreement Number W911NF-22-2-0037. The views and conclusions contained in this document are those of the authors and should not be interpreted as representing the official policies, either expressed or implied, of the Army Research Office or the U.S. Government. The U.S. Government is authorized to reproduce and distribute reprints for Government purposes notwithstanding any copyright notation herein.
 Sandia National Laboratories is a multimission laboratory managed and operated by National Technology and Engineering Solutions of Sandia, LLC, a wholly owned subsidiary of Honeywell International Inc., for the U.S. Department of Energy’s National Nuclear Security Administration under contract DE-NA0003525.

\appendix
\section{Effective $T_2^*$ from the autocorrelation function}
\label{appendix:T2star}

We can relate the time-dependent $T_2^*$ from measurements to the autocorrelation functions of the Overhauser field.  Our definition of this $T_2^*(t_{\rm avg})$ is not the current standard but more appropriately related to autocorrelation functions.  We obtain decay fits from averaging multiple Ramsey experiments that measure state return probabilities versus a waiting time $\tau$, but rather than averaging the associated $T_2^*$ quantities that characterize the $\exp{\left(\left(-\tau/T_2^*\right)^2\right)}$ decays, we average the $\left(1/T_2^*\right)^2$ quantities instead.  That is, we average decay rates in quadrature to get an expected decay rate in correspondence with $T_2^*(t_{\rm avg})$.

In the $\tau \rightarrow 0$ limit, no electron spin rotation occurs and the return probability is unity up to a visibility loss due to state preparation and measurement imperfections.  As a function of $\tau$, the return probability will oscillate according to an average rotation speed and decay according to the variance of the rotation speed.  In a double-dot experiment, the average rotation speed may be dominated by differences of the spin-orbit interaction of the two occupied dots~\cite{Jock2018} but the decay is typically dominated by the dynamics of the nuclear spin bath.  

For simplicity, assume a quasi-static Overhauser field which is constant within each Ramsey experiment on the timescale of $\tau$ but drifts from one Ramsey experiment to the next on the timescale of $t$.  
In a double dot experiment, these rotations refer to singlet/triplet rotations; in a single dot experiment, these would refer to $z$ rotations in the Bloch sphere.
In either case, the return probability as a function of  $\tau$ averaged over $N$ measurements is
\begin{equation}
P_{\rm return} = \frac{1}{N} \sum_{n=1}^{N} \cos^2\left(\Omega_n \tau / 2 \right).
\label{eq:return_probability}
\end{equation}
In the limit of large $N$, we can assume $\Omega_n$ is normally distributed (with a Gaussian probability density function) by the central limit theorem.  The $\Omega_n$ quantities are angular frequencies with implicit radian units.  We make the large $N$ approximation here but acknowledge that $N$ isn't always very large which leads to uncertainties.  This distribution has a mean and variance given respectively by
\begin{eqnarray}
\mu_{\Omega} &=& \frac{1}{N} \sum_{n=1}^{N} \Omega_n \\
\label{eq:Omega_variance}
\sigma_{\Omega}^2 &=& \frac{1}{N} \sum_{n=1}^{N} \left(\Omega_n - \mu_{\Omega}\right)^2.
\end{eqnarray}
Having defined these quantities, we substitute the probability density into and approximate the sum as an integral to obtain
\begin{equation}
P_{\rm return} \approx \int_{-\infty}^{\infty}
\frac{dx}{\sqrt{2 \pi \sigma_{\Omega}^2}}~ \mathrm{e}^{-x^2 / 2 \sigma_{\Omega}^2}
\cos^2\left(\left(x + \mu_{\Omega}\right) \tau /2\right).
\end{equation}
By trigonometric identities, 
\begin{eqnarray}
\nonumber
& \cos^2{\left(\left(x + \mu_{\Omega}\right) \tau / 2 \right)} = \frac{\cos{\left(\left(x + \mu_{\Omega}\right) \tau\right)} + 1}{2} \\
&= \frac{\cos{\left( x \tau \right)} 
\cos{\left( \mu_{\Omega} \tau \right)} - 
\sin{\left( x \tau \right)} 
\sin{\left( \mu_{\Omega} \tau \right)} + 1}{2}.
\end{eqnarray}
Since $\mathrm{e}^{-x^2 / 2 \sigma_{\Omega}^2} \sin{\left(x \tau\right)}$ is an odd function of $x$, the second term in the numerator above integrates to zero and we can drop it to leave
\begin{eqnarray}
\nonumber
P_{\rm return} &\approx& \int_{-\infty}^{\infty}
\frac{dx}{\sqrt{2 \pi \sigma_{\Omega}^2}}~ \mathrm{e}^{-x^2 / 2 \sigma_{\Omega}^2}
\frac{\cos{\left(\mu_{\Omega} \tau \right)} \cos{\left(x \tau \right)}+1}{2} \\
&=& \frac{\cos{\left(\mu_{\Omega} \tau \right)}
\mathrm{e}^{-\sigma_{\Omega}^2 \tau^2 / 2} + 1}{2},
\end{eqnarray}
by well-known integrals over normal distributions.

For this set of $N$ experimental measurement with Ramsey waiting time $\tau$, the decay is $\mathrm{e}^{-\sigma_{\Omega}^2 \tau^2 / 2} \equiv
\mathrm{e}^{(-\left(\tau/\hat{T}_2^*\right)^2)}$
so that $\hat{T}_2^* \approx \sqrt{2 / \sigma_{\Omega}^2}$ is the decoherence time we attribute to that set of measurements.  For a collection of such experiments, each collected over a duration of time $t$ with even spacing, we can average the Overhauser variances to determine the effective $T_2^*(t_{\rm avg}) \equiv
\sqrt{2 / \langle \sigma_{\Omega}^2 \rangle_t} = \sqrt{1 / \langle \left( 1 / T_2^* \right)^2 \rangle}$.  Approximating the summation in Eq.~(\ref{eq:Omega_variance}) with an integral and averaging, we obtain
\begin{eqnarray}
\langle \sigma_{\Omega}^2 \rangle_t &=& \left \langle \frac{1}{t} \int_{0}^{t} d t_1 \left(\Omega(t_1) - \frac{1}{t} \int_{0}^{t} d t_2 \Omega(t_2) \right)^2 \right \rangle \\
\nonumber
&=& \frac{1}{t} \int_{0}^{t} d t' 
\left \langle \Omega(t') \Omega(t') \right \rangle \\
\label{eq:avg_Omega_variance}
&& {} - \frac{1}{t^2} \int_{0}^{t} d t_1 \int_{0}^{t} d t_2 \left \langle \Omega(t_1) \Omega(t_2) \right \rangle.
\end{eqnarray}
In what follows, we invoke the wide-sense stationary property of the autocorrelation of $\Omega$, $\left \langle \Omega(t_1) \Omega(t_2) \right \rangle$, which results from our definition of the system dynamics as alternating periodically between two time-independent Hamiltonians (with and without the HFI as in Eq.~\ref{eq:rho_prime}). While one should think of the nature of the autocorrelation function as stroboscopic, meaning it evolves in discrete intervals $|t_1 - t_2|$ over multiples of the period $T_P = T_{\rm on} + T_{\rm off}$, for convenience in notation we disregard this stroboscopic nature in writing time integrals.
 Keep in mind that the integrals in the following steps should actually be effected as summations over discrete time points at intervals of $T_P$.  In actuality, the experiments we compare to in this paper are not strictly periodic in time, so this is all an approximation in any case.
 
To simplify the double integration in Eq.~(\ref{eq:avg_Omega_variance}), we transform the $t_1$, $t_2$ coordinates to $t_{\Delta} = (t_2 - t_1)$ and $t_{\Sigma} = (t_1 + t_2)$, yielding a Jacobian matrix
\begin{equation}
\left[
\begin{array}{cc}
\frac{\partial t_1}{\partial t_{\Sigma}} & \frac{\partial t_1}{\partial t_{\Delta}} \\
\frac{\partial t_2}{\partial t_{\Sigma}} & \frac{\partial t_2}{\partial t_{\Delta}} \\
\end{array}
\right] =
\left[
\begin{array}{cc}
\frac{1}{2} & -\frac{1}{2} \\
\frac{1}{2} & \frac{1}{2} \\
\end{array}
\right]
\end{equation}
with a determinant of $1/2$.  Thus,
\begin{eqnarray}
&& \int_0^t dt_1~ \int_0^t dt_2~ \langle \Omega(t_1) \Omega(t_2) \rangle \\
\nonumber
&=& \frac{1}{2} \int_{-t}^{t} dt_{\Delta}~ \int_{|t_{\Delta}|}^{2 t - |t_{\Delta}|} dt_{\Sigma}~ \left\langle \Omega\left(\frac{t_{\Sigma} - t_{\Delta}}{2}\right) \Omega\left(\frac{t_{\Sigma} + t_{\Delta}}{2}\right) \right\rangle \\
&=& \frac{1}{2} \int_{-t}^{t} dt_{\Delta}~ \int_{|t_{\Delta}|}^{2 t - |t_{\Delta}|} dt_{\Sigma}~ \langle \Omega(t_{\Delta}) \Omega(0) \rangle \\
&=& \int_{0}^{t} dt_{\Delta}~ \int_{t_{\Delta}}^{2 t - t_{\Delta}} dt_{\Sigma}~ \langle \Omega(t_{\Delta}) \Omega(0) \rangle \\
 &=& 2 \int_0^{t} dt'~ (t - t') \langle \Omega(t') \Omega(0) \rangle,
\end{eqnarray}
using the wide-sense stationary property $\left \langle \Omega(t_1) \Omega(t_2) \right \rangle = \left \langle \Omega(t_0 + |t_2 - t_1|) \Omega(t_0) \right \rangle$, leveraging the symmetry of the autocorrelation function, and renaming $t_{\Delta}$ as $t'$ in the last step.  Now we have, from Eq.~(\ref{eq:avg_Omega_variance}),
\begin{equation}
\label{eq:avg_Omega_variance2}
\langle \sigma_{\Omega}^2 \rangle_t = \left \langle \Omega(0) \Omega(0) \right \rangle
- \frac{2}{t^2} \int_0^{t} dt'~ (t - t') \langle \Omega(t') \Omega(0) \rangle.
 \end{equation}
 Finally, then, the effective $T_2^*(t_{\rm avg}) \equiv \sqrt{2 / \langle \sigma_{\Omega}^2 \rangle_t}$ in this approximation is
 \begin{equation}
 \label{eq:T2star_vs_t}
T_2^*(t_{\rm avg}) \approx \sqrt{\frac{2 t^2}{
t^2 \left \langle \Omega(0) \Omega(0) \right \rangle
- 2 \int_0^{t} dt'~ (t - t') \langle \Omega(t') \Omega(0) \rangle
}}.
 \end{equation}

If we take the $t \rightarrow \infty$ limit and assume that $\langle \Omega(0) \Omega(t') \rangle \rightarrow 0$ as $t' \rightarrow \infty$, then the second term in the denominator becomes negligible relative to the first and we get the ergodic $T_2^*$,
\begin{equation}
\label{eq:T2star}
T_2^* = \sqrt{\frac{2}{\left \langle \Omega(0) \Omega(0) \right \rangle}}.
\end{equation}
where $\Omega(t)$ remains an angular frequency with an implicit radian unit and $T_2^*$ has time units.
If we assume ergodicity and an infinite temperature nuclear spin bath, 
since typical operating temperatures ($\sim 100$~mK) are high relative to nuclear Zeeman energies ($\sim 1$~nK / mT), then this simply reduces to
\begin{eqnarray}
\label{eq:T2star_simplified}
T_2^* &=& \sqrt{\frac{2}{\sum_n A_n \left \langle \left(\hat{I}_{n}^{z} \right)^2 \right \rangle}}, \\
\left \langle \left(\hat{I}_{n}^{z} \right)^2 \right \rangle &=& \sum_{k=-2 I_n}^{2 I_n} \left(\frac{k}{2}\right)^2.
\end{eqnarray}
where $A_n$ is the HFI, $\hat{I}_n$ is the spin operator, and $I_n$ is the spin quantum number of the $n$th nuclear spin.  Note that cross-terms were excluded since they average to zero.

The question of drift compensation is a different matter.
Consider a dephasing error of a qubit accumulating over a time $\tau$ due 
to an unknown change of the Overhauser rotation rate $\Delta \Omega$ that has drifted away from its previously known measurement over a time $\Delta t$.  
After a projective measurement most directly sensitive to this dephasing (e.g., from quantum error correction measurements), the contribution to the error due to this drift is
\begin{eqnarray}
P_{\rm err} &=& 1 - \left \langle \cos^2{\left(\Delta \Omega \tau / 2 \right)} \right \rangle \\
&\approx& \frac{1 - \mathrm{e}^{-\sigma_{\Delta \Omega}^2 \tau^2 / 2}}{2}
\end{eqnarray}
where the variance of $\Delta \Omega$, $\sigma_{\Delta \Omega}^2$, is given by
\begin{eqnarray}
\sigma_{\Delta \Omega}^2 &=& \left \langle \left( \Omega(\Delta t) - \Omega(0) \right)^2 \right \rangle \\
&=& 2 \left \langle \Omega(0) \Omega(0) \right \rangle - 2 \left \langle \Omega(\Delta t) \Omega(0) \right \rangle
\end{eqnarray}
under the assumption that $\Omega(\Delta t)$ is normally distributed with a mean of $\Omega(0)$, the previously known Overhauser rotation, and under the wide-sense stationary approximation once again.
We now define a drift-compensated, effective $\tilde{T}_2^*(\Delta t)$ as
\begin{equation}
\label{eq:drift_compensated_T2star_def}
\tilde{T}_2^*(\Delta t) = \sqrt{\frac{2}{ \sigma_{\Delta  \Omega}^2}} = \frac{1}{\sqrt{\left \langle \Omega(0) \Omega(0) \right \rangle - \left \langle \Omega(\Delta t) \Omega(0) \right \rangle}}
\end{equation}
such that
\begin{equation}
P_{\rm err} \approx \frac{1 - \mathrm{e}^{-
\left(\tau / \tilde{T}_2^*(\Delta t)\right)^2}}{2}.
\end{equation}

Notice that in the limit of $\Delta t \rightarrow \infty$, the variance with drift compensation is twice the variance without drift compensation and, correspondingly, $\tilde{T}_2^*(\Delta t \rightarrow \infty) = T_2^*/\sqrt{2}$ [comparing with the ergodic $T_2^*$ of Eq.~(\ref{eq:T2star})].  This seems counter-intuitive, but it is easily understood if you consider that drift compensation in the long time limit represents the subtraction of two independent, random variables with a variance that is the sum of the variance of each one.  That is, compensating for something one has no real knowledge about is worse than no compensation at all.

\section{Overhauser dynamics as classical noise and its PSD}
\label{appendix:classical_PSD}

As an alternative to an expensive quantum description of nuclear bath flip-flop dynamics, 
observables manifesting from a sufficiently large bath can be approximated within a classical noise formalism. 
An ensemble of TLFs is conveniently described as a function of a discrete stochastic variable, $\chi$. 
Let us consider here TLFs composed of two coupled $I=1/2$ nuclear-spin quanta, which are governed by
\begin{equation}
f(\chi): \begin{cases}
    x_+ \rightarrow (|\uparrow\downarrow\rangle + |\downarrow\uparrow \rangle)\\
    x_- \rightarrow (|\uparrow\downarrow\rangle - |\downarrow\uparrow \rangle).
  \end{cases}
\end{equation}
which has an associated probability distribution function, $P_{\chi}(x)$, that is both normalized and symmetrically-distributed
over the domain $\chi \in \{x_+,x_-\}$, i.e., $P_{\chi}(x_+) = P_{\chi}(x_-) = 1/2$.

A given TLF noise source is assumed to manifest temporally as a telegraph process, $\eta(t)$, characterized by Markovian, continuous-time stochastic behavior. Upon specifying an initial state and a period, the system can advance through one among $i$ possible deterministic trajectories, $\eta_i(t)$, until it arrives at a common final state. The family of trajectories is characterized by an event-rate $\omega$ and, invoking the Chapman-Kolmogorov equation, its $k^{\mathrm th}$ joint probability distribution is
\begin{multline}
P_{\eta}^{(k)}(\eta_1, t_1;\ldots;\eta_k,t_k)= \nonumber \\ 
\frac{1}{2}P_{\eta}^{(1)}(\eta_k,t_k) \prod_{j=1}^{k-1} \left(1+\eta_j\eta_{j+1} \exp{(-2\omega(t_j-t_{j+1}))} \right)
\end{multline}
for $\eta_j = \pm 1$. The PDF after a single event is
\begin{equation}
P_{\eta}^{(1)} (\eta_1,t_1) = \frac{1}{2} + \frac{p \eta_1 \exp{(-\gamma t)}}{2}
\label{eq:Toby}
\end{equation}
with $p \in {-1,1}$ and $\gamma=2\omega$. 

To obtain the PSD of the TLF ensemble, one may FT the autocorrelation function, 
$C(t_1,t_2)\equiv \langle\eta(t_1)\eta(t_2)$. For TLFs, the corresponding expression,
\begin{align} 
\langle\eta(t_1)\eta(t_2)\rangle &= \sum_{\eta_1,\eta_2=\pm 1} \eta_1\eta_2 P_{\eta}^{(2)} (\eta_1,t_1;\eta_2,t_2) \nonumber \\
&=\exp{(-\gamma(t_1-t_2))} \sum_{\eta_2 = \pm 1} \eta_2^2 P_{\eta}^(1) (\eta_2,t_2) \nonumber \\
&= \exp(-\gamma(t_1 - t_2)),
\label{classical_corrfn}
\end{align}
may be shifted freely in time, since telegraph noise is a wide-sense stationary stochastic process, such that $C_{\eta}(t-t_0,t_0-t_0)=\langle \eta(t)\eta(0) \rangle$. Finally, invoking the Wiener-Khinchin theorem, the Fourier transform (FT) of Eq.\ \ref{classical_corrfn} is simply
\begin{equation}
S_{\eta} (\omega) = \frac{4\gamma}{(2\gamma)^2 + \omega^2}
\label{eq:classical_PSD}
\end{equation}
which exhibits a Lorentzian spectral line shape. 

By comparing the above to the analysis in Ref.\ \cite{Gorman2012}, two corrections should be applied to their Eq.~(3): a factor of $1/N$ is missing and the summation over $k$ should be inside the squaring operation. Referring the reader to Ref.\ \cite{Gorman2012} for the full details, we will summarize their analysis, which is generally applicable to an ensemble multi-level fluctuators (MLFs). For numerical convenience, they transformed Eq.\ \ref{classical_corrfn} into a matrix-vector form
\begin{align}
\langle \eta(t) \eta(0) \rangle &= \sum_{i,j} P(\eta_j , t|\eta_i, 0) \eta_j P(\eta_i (0))\eta_i \nonumber \\
     &= \frac{1}{N} \sum_{i,j} \eta_i [\exp{(\bf{\Gamma}|t|)}]_{ij} \eta_j \nonumber \\
     &= \frac{1}{N} \vec{\eta}^{\dagger} {\bf{V}}^{\dagger} \exp{(\Lambda|t|)} \bf{V} \vec{\eta} \nonumber \\
     &= \vec{b} \exp{(\bf{\Lambda}|t|)} \vec{b},
\label{eq:gorman_corrfn}
\end{align}
with $\bf{V}$ and $\bf{\Lambda}$ the eigenvectors and eigenvalues of the real-symmetric rate matrix $\bf{\Gamma}$ governing the evolution of the TLFs, and where the transformed noise amplitude was introduced as $\vec{b} = \bf{V}\vec{\eta}/\sqrt{N}$. Taking the FT of Eq.\ \ref{eq:gorman_corrfn}, we arrive at
\begin{align}
S(\omega;\vec{\eta},\gamma) &= \int_{-\infty}^{\infty} C(t) \exp{(-i\omega t)} dt \nonumber \\
        &= \sum_{j} \frac{-2(\Sigma_k V_{j,k}\eta_k / \sqrt{N})^2 \lambda_j}{\lambda^2_j + \omega^2}.
\label{eq:gormanMLF}
\end{align}
This general expression for MLFs reduces to the specific case of TLFs --- for which $N=2$, $V_{j,k} = ((1,1),(-1,1))/\sqrt{2}$, and $\lambda = (0,-2\gamma)$, and recalling that $\vec{\eta}$ is a unit vector -- to arrive again at Eq.\ \ref{eq:classical_PSD}.

\section{Automatic determination of model dimensions}
\label{appendix:auto_cylinder}

This appendix presents automated methods for two successive constraints of the minimum geometrical dimensions required to achieve a desired accuracy. 
The objective is to determine the minimum dimensions of a volume
that contains all of the spins sufficient to calculate  $T_2^*(t_{\rm avg})$ within a given accuracy threshold.
The minimum dimensions are determined by the desired accuracy threshold, the dimensions of the quantum dot $\psi(\vec{r})$, concentrations of spinful nuclei, and the number of pairs per spin for generating the clusters used in the simulation.

In the first approach, we evaluate the normalized difference of ergodic $T_2^{\ast}$ values derived from volumes having infinite [$T_2^{\ast}(t\rightarrow\infty;V\rightarrow\infty)\equiv T_2^{\ast}(t_{\infty};V_{\infty})$] and finite [$T_2^{\ast}(t\rightarrow\infty;V\rightarrow V_0)\equiv T_2^{\ast}(t_{\infty};V_0)$] extents. For a system where the quantum dot spin interacts with spinful nuclei, the effective relaxation time can be expressed as $1/T_2^{\ast} \propto \sum_i A_i^2(\eta_i)$, where 
$A_i(\eta_i) \approx \eta_i^2 |\psi(R_i)|^2$ is the hyperfine strength of nucleus $i$ at nuclear position $R_i$ and with $\eta_i$ its bunching factor. Substituting definitions, we have
\begin{widetext}
\begin{equation}
\mathrm{threshold}   \geq \frac{T_2^{\ast}(t_{\infty};V_{\infty}) - T_2^{\ast}(t_{\infty};V_0)}{
                    T_2^{\ast}(t_{\infty};V_{\infty})  }
                    \gtrapprox \frac{ \left[\int_{0}^{\infty} \psi^4(r) \rho(r;\eta^2) d^3r\right]^{-0.5} 
                    -  \left[\int_{0}^{V_0} \psi^4(r) \rho(r;\eta^2)d^3r\right]^{-0.5}  
                    }{  \left[\int_{0}^{\infty} \psi^4(r) \rho(r; \eta^2)d^3r\right]^{-0.5} },
\label{eq:T2star_residual_sphere}
\end{equation}
\end{widetext}
where the latter inequality indicates that we are approximating the summations over discrete nuclear positions using integrals
with $\rho(\vec{r}; \eta^2)$ is the density profile of nuclear spins multiplied by the appropriate $\eta^2$.  When the spin density functions have multiple components, they take the form $\rho(x) = \sum_i \rho_i(x) \eta_i^2 $, where each spinful nucleus is assigned its own profile and bunching factor.

For the specific case of an elliptic cylinder with height ($h$) and width dimensions ($r_x$ and $r_y$), Eq.\ \ref{eq:T2star_residual_sphere} is separable into a pair of independent expressions which are soluble for the optimal vertical ($r_{0,x}$ and $r_{0,y}$) and horizontal ($h_0$) dimensions: 
\begin{widetext}
\begin{align}
\mathrm{threshold}  &\gtrapprox  \frac{ \left[\int_{0}^{\infty}\psi^4(z) \rho(z; \eta^2)dz\right]^{-0.5}  -  \left[\int_{0}^{h_0}\psi^4(r) \rho(z; \eta^2)dz\right]^{-0.5}  }{  \left[ \int_{0}^{\infty}\psi^4(z) \rho(z;\eta^2)dz\right]^{-0.5}  }, \nonumber \\
\mathrm{threshold}  &\gtrapprox  \frac{  \left[\int_{0}^{\infty}\int_{0}^{\infty}\psi^4(r_x,r_y) \rho(r_x,r_y;\eta^2)dr_x dr_y\right]^{-0.5}  -  \left[\int_{0}^{r_{0,x}}\int_{0}^{r_{0,y}}\psi^4(r_x,r_y) \rho(r_x,r_y;\eta^2)dr_x dr_y\right]^{-0.5}  }{  \left[\int_{0}^{\infty} \int_{0}^{\infty}\psi^4(r_x,r_y) \rho(r_x,r_y; \eta^2)dr_x dr_y\right]^{-0.5} },
\label{eq:T2star_residual_cylinder}
\end{align}
\end{widetext}
where the optimal quantities correspond to $h_0$ and $r_0$ in Fig.\ \ref{fig:cyl_cutoff}.  Integration can be performed analytically or numerically, depending on the form of the quantum-dot wave function.

\begin{figure}[t]
\includegraphics[width=.9\columnwidth]{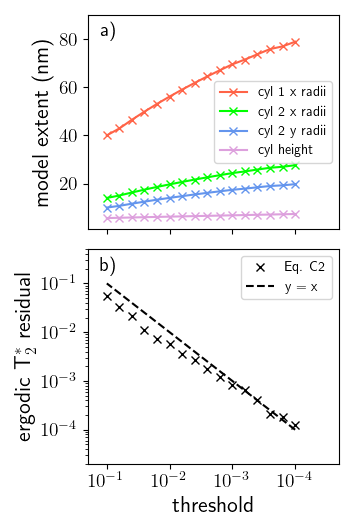}
\caption{Threshold convergence studies involving (a) the model space dimensions, in this case the height and radii of two elliptic cylinders containing each quantum dot, and (b) the standard ergodic limit $T_2^{\ast}(t=\infty)$ residual, defined as the relative error compared with reference values generated using Eq.\ \ref{eq:T2star_residual_cylinder} with the infinite integrals generated numerically with an effectively converged threshold value of 1$\times 10^{-6}$ with 3 pairs per spin. The dashed line corresponds to $x=y$ and residual values are shown as symbols. The underlying double-dot model system corresponds to the Foster entry in Table \ref{tab:parameters}. }
\label{fig:thresh_conv1}
\end{figure}

Fig.~\ref{fig:thresh_conv1} presents convergence tests conducted to validate our computer implementation of this first method for automatically determining the geometrical extents. The first panel depicts the dependence of the model dimensions -- comprising the radii and heights of two elliptic cylinders containing disjoint spin sets -- on the user-defined threshold value. For clarity, we note that the y-radius plot for cylinder 1 has been omitted from panel (a) because both cylinders have equivalent y radial extents. The results reveal a non-linear relationship between the model parameters and the threshold, suggesting that the cylinder dimensions would be overestimated by an {\it ad hoc} numerical extrapolation (e.g., a linear fit on a log-linear scale). Meanwhile, the second panel demonstrates the quasi-linear relationship between user-defined thresholds and the ergodic $T_2^{\ast}$ residuals. This aligns well with our expectations based on the form of Eq.~\ref{eq:T2star_residual_cylinder}. 

\begin{figure}[t]
\includegraphics[width=.9\columnwidth]{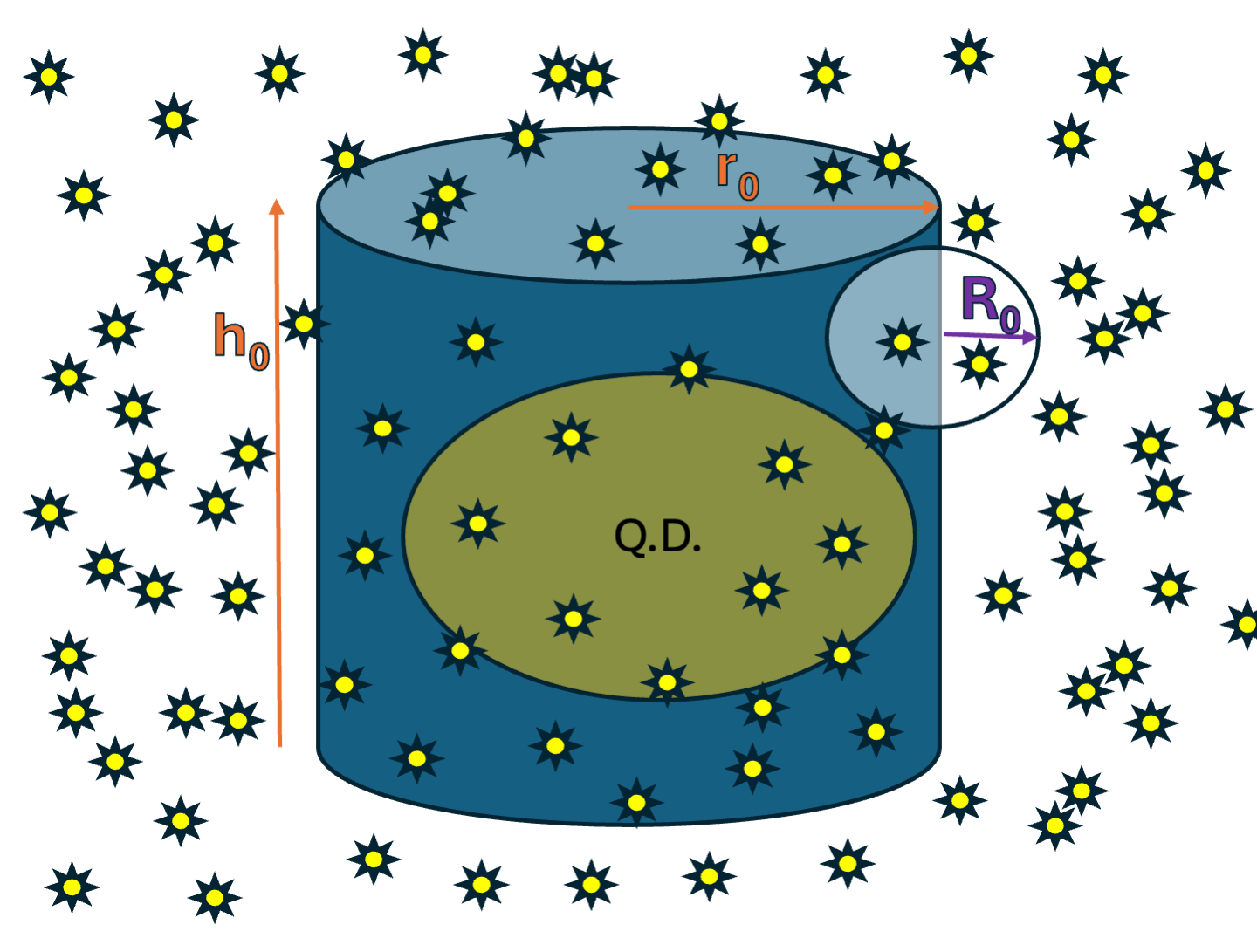}
\caption{Cartoon depiction of a three-dimensional spin bath immersed in a spheroidal quantum dot (yellow oval).
         Dimensions for the blue cylinder ($h_0, r_0$) are found using Eq.\ \ref{eq:T2star_residual_cylinder}, while Eq.\ \ref{eq:radius} provides an analytical expresssion for the radius ($R_0$) 
         of the transparent sphere. The transparent sphere represents an extension of the cylinder necessary for including fringe spins that may be important in dynamical simulations.}
\label{fig:cyl_cutoff}
\end{figure}

\begin{figure}[t]
\includegraphics[width=.9\columnwidth]{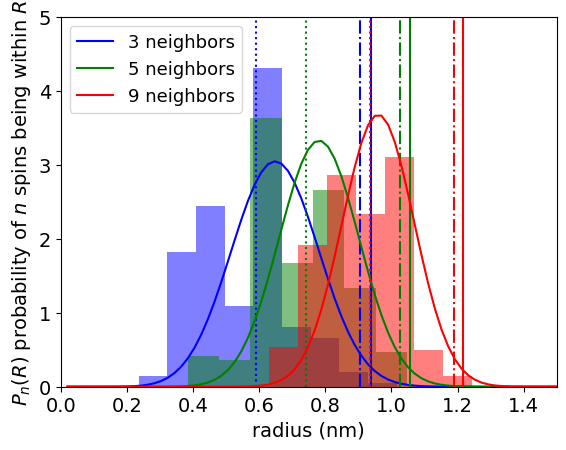}
\caption{A comparison of analytic (curves) and numerical (histograms) distribution functions corresponding to the number of nearest-neighbor $^{29}$Si nuclei within a sphere centered within a volume of natural silicon. Vertical lines mark the radii corresponding to threshold values of 0.5 (dotted), 0.01 (dot-dashed), and 0.005 (solid).}
\label{fig:rad_probs}
\end{figure}

While this first method is sufficient for estimating ergodic $T_2^*$ values,  we need to extend the simulation domain further for well-justified dynamical simulations (e.g., to obtain $T_2^*(t_{\rm avg})$, autocorrelation functions, or PSDs) because of the potential for nuclear spins within the domain of the first method to interact (e.g., flip-flop) with nuclear spins outside of that domain. To illustrate the situation, consider a sphere with radius $R_0$ centered at the edge of the cylinder as in Fig.\ \ref{fig:cyl_cutoff}. In this scenario, two spinful nuclei are encompassed within the sphere centered around the radius edge. Given that such spin pairs can contribute a relatively strong dipolar interaction, they must be included in the dynamics. These considerations must also be generalized to $k$ spins within the sphere.

For a random distribution of spin impurities at a specific concentration, we can derive the associated probability density function for nearest neighbors, next-nearest neighbors, and so on. For example, a wafer of elemental silicon having a natural isotopic abundance of $^{29}$Si, a spherical volume $V(R_0)$ expanded around a given position will encompass $P$(Si$^{29}) [N \approx \rho_{\mathrm Si} - 1]$ total spinful sites, with $P$(Si$^{29})$ being the isotopic enrichment. 
The probability of finding $X$ spins within $V$, assuming $N$ independent trials, can be expressed as a binomial distribution \cite{Hall2014}:
\begin{eqnarray}
P(X|N, P(^{29}\mathrm{Si})) &\approx& \frac{(V/V_0)!}{X!(V/V_0 - X)!} P(^{29}\mathrm{Si})^X \nonumber \\
       & \times &  (1 - P(^{29}\mathrm{Si}))^{V/V_0 - X},
\label{eq:binomial}
\end{eqnarray}
with $V_0 = P(^{29}\mathrm{Si}) / \bar{\rho}(^{29}\mathrm{Si})$ defined in terms of $\bar{\rho}(\mathrm{Si}^{29})$, the average spinful impurity density. As isotopic enrichment increases, this binomial distribution approaches a Poisson distribution:
\begin{equation}
P(X|N, P({\mathrm Si}^{29})) 
    \approx \frac{1}{X!} (\xi r^3 )^X \exp(-\xi r^3)
\label{eq:poisson}
\end{equation}
where $\xi \equiv 4/3\ \pi \bar{\rho}({\mathrm Si}^{29})$. As was done for $^{29}$Si, this analysis may be applied to any spinful component.  

To find the probability that $k$ other spins are within a sphere with volume $V(R_0)$, consider two concentric spheres with radii $r_{k-1}$ and $r_{k}$ encompassing $k-1$ and $k$ spins, respectively. The PDF for the distance to the $k$th impurity is given by:
\begin{equation}
P(r_k) = 3 \xi r_{k}^2 \exp{-\xi(r_k^3 - r_{k-1}^3)}.
\label{eq:pdf_k}
\end{equation}
The joint PDF is 
\begin{eqnarray}
P(r_1 \ldots r_k) &=& \prod_{j=1}^k p_r (r_j) \nonumber \\
                  &=& (3 \xi)^k r_1^2 \cdots r_k^2 \exp(- \xi r_k^3),
\label{eq:jpdf_k}
\end{eqnarray}
where we have set $r_{0}=0$. The probability of finding the $k$th spin at a distance $r_k$ is given by
\begin{equation}
P_k(r_k) = \frac{4\pi n r_k^2}{(k-1)!} \left( \frac{4\pi n r_k^3}{3}\right)^{k-1} \exp{\left(-\frac{4\pi n r_k^3}{3}\right)}.
\label{eq:prob_k}
\end{equation}
Integrating the PDF gives the probability of the $k$th spin existing on the perimeter of $V(R_0)$, leading to an analytical expression for the radial cutoff:
\begin{equation}
R_0 = \left[ \frac{3}{4\pi n} \Gamma^{-1}(k-\frac{2}{3}, 1-\mathrm{threshold}) \right]^{\frac{1}{3}}   
\label{eq:radius}
\end{equation}
where $n = \chi/V_0$ with $\chi$ the fraction of target nuclei.

Fig.\ \ref{fig:rad_probs} provides validation of our derivation and implementation of Eq.\ \ref{eq:prob_k}. Histograms were generated binning the the $k$-th nearest-neighbor was placed within 10,000 random instances of $^{29}$Si spin distributions. Good agreement between numerical and analytical results were found when the distributions returned by Eq.\ \ref{eq:prob_k} for $k$ = 3, 5, and 9 were overlaid on the histograms. For $k<3$, the sparsity of available lattice sites interfered with the statistics. Agreement between analytic and numerical distributions was observed to improve with increasing $k$. The convergence of radial values with respect to input threshold values is also demonstrated in Fig.\ \ref{fig:rad_probs}.

\begin{figure}[t]
\includegraphics[width=\columnwidth]{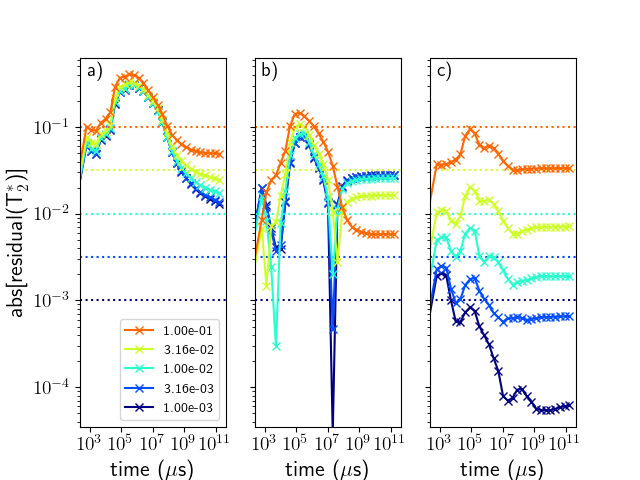}
\caption{Convergence of the effective $T_2^*(t_{\rm avg})$ with respect to the threshold parameter and the number of pairs per spin (nps), with nps set as 1, 2, and 3 in panels (a), (b), and (c), respectively. Rather than only taking ergodic values in the residual formula, as defined in Fig.~\ref{fig:thresh_conv1}, here all values entering the residual formula correspond to the indicated time. Dotted lines mark threshold values and they share the color of corresponding simulations.}
\label{fig:thresh_conv2}
\end{figure}

In Fig.~\ref{fig:thresh_conv2} we present a study of $T_2^{\ast}(t)$ convergence properties considering both the threshold value and the number of pairs per spin (nps). Reference values used in the residual calculations were generated with a threshold value of 1$\times 10^6$ and nps set to three. Focusing first on the results collected for nps = 3 in Fig.~\ref{fig:thresh_conv2}(c), the outcomes align with expectations. The residuals predominantly remain below the corresponding threshold value, consistent with the findings in Fig.~\ref{fig:thresh_conv1}. The few exceptions occur at times for which infinities when infinities had to be replaced with a large number (1$\times 10^{24}$) to facilitate valid numerical operations. 

Comparing panels (a)–(c) in Fig.\ \ref{fig:thresh_conv2}, where nps was set to 1, 2, and 3, respectively, it is evident that the simulation accuracy is strongly dependent on the value of nps. This is due to the dual role that nps plays in our spin-dynamics implementation. First, it determines the number of partners for each spin when assigning spin pairs with the strongest dipolar interactions. Second, it expands the cylinder dimensions according to the value given by Eq.~\ref{eq:radius}. 
The importance of the former effect can be seen by comparing subplots (a), (b), and (c) in Fig.~\ref{fig:thresh_conv2}.  For example, the relative errors of Fig.\ \ref{fig:thresh_conv2}(a) appear to saturate in the small threshold limit indicating a limit to what can be achieved when only 1 pair per spin is included in the cluster calculation.  In contrast, Fig.\ \ref{fig:thresh_conv2}(c) achieves much smaller relative errors by including 3 pairs per spin.

\pagebreak
\bibliography{bibliography}
\end{document}